\documentstyle[epsf]{mn}
\def\epsfsize#1#2{\hsize}

 \title{Evolutionary synthesis of stellar populations:\\* a modular tool}
\author[C. Maraston]
        {Claudia Maraston \\
         Department of Astronomy, University of Bologna, via Zamboni 33,
         40126 Bologna, Italy}
\date {Accepted Received ; in original form }

\pubyear{1997}

\begin{document}

\maketitle

\begin{abstract}
A new tool for the Evolutionary Synthesis of Stellar Populations
 (EPS) is presented, which is based on three independent matrices, giving respectively: 1) the fuel 
consumption during each evolutionary phase as a function of stellar mass; 
2) the typical temperatures and gravities during such phases; 3)  
colours and bolometric corrections as functions of gravity and temperature. 
 The modular structure of the code allows to easily assess the impact on the 
synthetic spectral energy distribution of the various assumptions and model 
ingredients, such as, for example, uncertainties in stellar evolutionary 
models, mixing length, the temperature distribution of horizontal branch (HB) 
stars, AGB mass loss, and colour-temperature transformations.
The so-called ``AGB-Phase Transition'' in Magellanic Cloud clusters is used to calibrate the contribution of the
Thermally Pulsing Asymptotic Giant Branch phase (TP-AGB) to the synthetic 
integrated luminosity.
As an illustrative example, solar metallicity ($Y=0.27,Z=0.02$) models, with 
ages ranging between 30 {\rm Myr} and 15 {\rm Gyr} and various choices for the 
slope of the Initial Mass Function (IMF), are presented. 
Synthetic broad band colours and the luminosity contributions of the various 
evolutionary stages are compared with LMC and Galactic globular cluster data.
In all these cases, a good agreement is found.
Finally, we show the evolution of stellar mass-to-light ratios in 
the bolometric and $U$,$B$,$V$,$R$, and $K$ passbands, in which the 
contribution of stellar remnants is accounted for. 

\end{abstract}


\section{Introduction}

In order to study the formation and the subsequent evolution of galaxies, we 
have to understand the overall properties of their stellar content. In a 
galaxy, a mixture of stellar populations of different ages and chemical 
compositions are present and many efforts have been made in recent years, 
inspired by the pioneer work of Tinsley (1980), aimed at modelling the 
spectral evolution of the various morphological types, from ellipticals to 
spirals (e.g. Bruzual 1983; Arimoto \& Yoshii 1987; 
Guiderdoni \& Rocca-Volmerange 1987; Bruzual \& Charlot 1993; 
Bressan, Chiosi \& Fagotto 1994; Tantalo {\rm et al.} 1996). 

The building blocks of Evolutionary Population Synthesis (EPS) are
models for Simple Stellar Populations (SSPs), that are 
assemblies of chemically homogeneous and coeval single stars. 
Thus, before facing the problem of modelling a complex stellar population, as a
 galaxy, the preliminary condition is to check the accuracy of SSPs computations
  and their adequacy to reproduce the observable features of stellar clusters, 
that better resembling the definition of SSP.
Two main approaches have been followed in computing EPS for SSPs. 
The `Isochrone Synthesis' technique (Charlot \& Bruzual 1991) consists 
in summing up the contributions to the flux in the various passbands of all 
mass-bins along one isochrone, after assuming an Initial Mass Function (IMF). 
The integration starts from a lower mass limit and ends at the latest 
mass point on the isochrone itself, that usually coincides with the end of the 
so-called Early Asymptotic Giant Branch (E-AGB) phase. 
Later stellar phases are then added following individual recipes (e.g. Charlot 
\& Bruzual 1991). 
An alternative approach to compute the luminosity contributions of Post-Main 
Sequence evolutionary stages is based on the so-called Fuel Consumption 
theorem (FCT, Renzini \& Buzzoni 1986).
Here the main ingredient of synthesis is the amount of nuclear fuel 
(i.e. the hydrogen and/or helium mass) that is burned in each evolutionary 
stage. The fuel, a natural product of evolutionary stellar sequences, is 
converted into bolometric luminosity thanks to the FCT. The luminosities in the
 various passbands are then computed thanks to a temperature/gravity-colour set 
 of transformations. 
In principle the two approaches are equivalent, as both are based on 
stellar evolutionary models, but their practical implementations may result in
 sizable differences. In the isochrone synthesis method, the integration 
 variable is the stellar mass, that varies along the isochrone. So, bright, 
 but short-lived evolutionary stages span a very narrow range in stellar mass 
 and integrations must use very small $\Delta M$ steps. (We have noticed that 
 mass steps as small as $10^{-5}$ $M_{\odot}$ are required to achieve a few 
 per cent stability in numerical results.) Coarse mass zoning and poor 
 interpolation algorithms are responsible for the sizable fluctuations 
 occasionally exhibited by population models constructed with the isochrone 
 synthesis method (see e.g. Fig. 7 in Charlot \& Bruzual 1991). The method 
 based on the FCT is instead much more robust in this respect, as the 
 integration variable, the fuel, is directly proportional to the contribution 
 of the various phases to the total luminosity.

At present this method has been followed by Buzzoni (1989) to construct SSPs 
for ages older than 4 {\rm Gyr}. In this work we extend the approach to young 
and intermediate age populations and present SSP models for solar metallicity 
and ages from 30 {\rm Myr} to 15 {\rm Gyr}. The synthesis computational code 
has been constructed with the aim of better controlling the influence of 
various model ingredients on SSP models. The comparison with Magellanic Clouds 
and Galactic globular clusters allows a detailed test for the results. 
The paper is organised as follows. In Section 2 a detailed description
of the algorithm is presented; Section 3 illustrates all
model ingredients: stellar input and temperature-colour transformations. 
Section 4 describes the results and Section 5 shows the various 
observational tests performed. Finally, in Section 6, our conclusions are drawn. 

\section[]{The Algorithm}

The EPS code used in this paper is based on three independent sets of matrices
containing the ingredients of the synthesis, namely:

\begin{enumerate} 
  \item {\it The energetics}: the nuclear fuel burned during each 
    evolutionary phase as a function of stellar mass; 
  \item {\it The surface parameters}: i.e. the effective temperatures
         and surface gravities during such phases; 
  \item {\it The transformations to observables}: i.e. colours and bolometric
             corrections as functions of gravity and temperature.
\end{enumerate}

Following Renzini \& Buzzoni (1986; hereafter RB86), the total bolometric 
luminosity of a Simple Stellar 
Population (SSP) of age $t$ can be written as the sum of two terms,
one for the core hydrogen burning stars (the Main Sequence (MS) stars), one for
 the evolved ones (the Post-Main Sequence (PMS) stars)

\begin{equation}
 L_{\rm T}^{\rm bol}(t) = L_{\rm MS}^{\rm bol}(t) + L_{\rm PMS}^{\rm bol}(t).
\end{equation}
$ L_{\rm MS}^{\rm bol} $ depends on the adopted mass-luminosity relation 
$ L(M,t) $ and Initial Mass Function (IMF) $\Psi(M)=A{M}^{-(1+x)}$
($A$ being the scale factor giving the size of the stellar population).
$ L_{\rm PMS}^{\rm bol}$ is directly connected with the total nuclear fuel 
burned by stars populating the various evolved stages, through the FCT (see 2.1). 

To evaluate the MS bolometric luminosity, we need a set of 
isochrones, on which to perform a numerical integration weigthed on the IMF
\begin{equation}
 L_{\rm MS}^{\rm bol}(t)=\int_{M_{\rm inf}}^{M_{\rm TO}(t)} L(M,t)\Psi(M){\rm
 dM}
\end{equation}
The lower integration limit is the minimum mass igniting {\rm H} in the core 
($ M_{\rm inf} $). $ M_{\rm TO}$ (t) (the turnoff mass) is the mass in the
verge of exhausting hydrogen in the core and leaving the MS. 
We chose as $ M_{\rm TO}$ (t) the model with the maximum effective temperature 
along the isochrone as, by definition, the turnoff point is the
bluest point on the Hertzprung-Russel (H-R) diagram. 
The relation between the age and the corresponding turnoff mass is the
{\it evolutionary clock} of a stellar population.

\subsection[]{Post-Main Sequence contributions: the Fuel Consumption Theorem}

The Fuel Consumption theorem (FCT) states that: 
{\it the contribution of stars in any given Post-MS stage to the integrated 
bolometric luminosity of a SSP is directly proportional to the amount of fuel
burned during that stage} (RB86).
Its analytical form is 
\begin{equation}
 L_{\rm j}^{\rm bol}(t) = 9.75\times 10^{10} b(t)F_{\rm j}(M_{\rm TO})~~~~~~~~~~~~~(L_{\sun}) 
\end{equation}
where $ L_{\rm j}^{\rm bol}(t) $ is the total bolometric luminosity 
of stars populating the PMS stage $ j $ at the age $t$. $b(t)$,
the ``{\it evolutionary flux}'', is the number of stars evolving off the Main 
Sequence per year. $ F_{\rm j}(M_{\rm TO})$ is the nuclear fuel (in $M_{\sun}$)
 burned by stars with $M$=$M_{\rm TO}$ in their {\rm j}-th evolutionary PMS 
 phase.

The function $b(t)$ is defined as
\begin{equation}
 b(t) = \Psi(M_{\rm TO}) |{{dM}_{\rm TO}\over{dt}}|~~~~~~~~~~~~~({\rm stars/yr})  
\end{equation}
where $\Psi(M_{\rm TO})$ is the IMF computed for $M=M_{\rm TO}$ and 
{\.M}${_{\rm TO}} $ is the time derivative of the relation $ M_{\rm TO}(t)$.
At any SSP age, the mass of evolved stars is only slightly 
greater than $M_{\rm TO}(t)$. Hence, to a fair approximation $b(t)$ also gives
 the rate at which stars enter or leave any specific post-MS evolutionary 
stage.

The fuel consumption $ F_{\rm j}$ is given by
\begin{equation}
F_{\rm j}={m_{\rm j}}^{\rm H} + 0.1{m_{\rm j}}^{\rm He} ~~~~~~~~~~~~~~~~(M_{\sun})  
\end{equation}
where $ {m_{\rm j}}^{\rm H} $ and ${m_{\rm j}}^{\rm He}$ are the amount of 
hydrogen and helium burned during the phase $j$, the coefficient 0.1 
taking into account that the energy release from helium-burning is ${\simeq 1/10}$ 
than from {\rm H}-burning.  
The close coincidence in mass between evolved stars and $M_{\rm TO}(t)$ 
allows us to refer the PMS fuels to the values proper to a mass equal to 
$M_{\rm TO}(t)$. 
As a consequence, mass and age indices are equivalent for PMS stars
and the index $j$ will be used to label star masses in the MS and evolutionary 
stages in the PMS.

This approximation, in which one assumes the flux $b(t)$ to be constant
through the various PMS stages, was extensively criticized by Charlot \&
Bruzual 1991. However, it has been demonstrated that this is actually a very
accurate approximation (Renzini 1994).

\subsection[]{From Bolometric to Monochromatic}

To compute the SSP total luminosity in the generic photometric
band $\lambda$, i.e. 
\begin{equation}
 L_{T}^{\lambda}(t) = L_{MS}^{\lambda}(t) + L_{PMS}^{\lambda}(t),
\end{equation}
we need temperature/gravity - bolometric correction/colour sets of 
transformations. 
The $ (T_{\rm e},g)_{\rm ij} $ matrix contains the effective temperatures 
and the surface gravities of stars in the $j$-th phase, at the SSP age 
$i$ (see 3.3 for a detailed description of adopted values). 
The colour index/bolometric correction matrix $ (CI,BC)_{\rm lm} $
contains the trasformations to the observables, i.e. bolometric
corrections and colours as functions of the surface parameters 
$ (T_{\rm e},g) $. $ l $ is an effective temperature index and $ m $
refers to surface gravity: we defer to 3.4 for a detailed description of 
adopted sets.
The ${\lambda}$-luminosity of the $j$-star at the SSP age $i$, 
$L_{\rm ij}^{\lambda}$, is a fraction of its bolometric luminosity 
$L_{\rm ij}^{\rm bol}$, as  

\begin{equation}
L_{\rm ij}^{\lambda}=Q_{\rm ij}^{\lambda}\cdot 
L_{\rm ij}^{\rm bol}~~~~~~~~~~~~~~~~~~~~~~~~~~~~~~~~~~~~~~~~({\rm
L_{\odot}^{\lambda}}) 
\end{equation}
Luminosities in eq. (7) are in solar units and $Q_{\rm ij}^{\lambda}$
is defined as
\begin{equation}
  Q_{\rm ij}^{\lambda}=
 10^{-0.4[{BC^{\rm \lambda}_{\sun}}-
 BC_{\rm ij}^{\lambda}]}
\end{equation}
where $BC_{\rm ij}^{\rm \lambda}$ is the bolometric correction to the 
$\lambda$-band and $BC^{\rm \lambda}_{\sun} $ the same for the Sun.

In order to get the luminosity contributions of PMS phases, 
we divide each phase $j$ into a certain number of photometric sub-phases,
burning an equal fraction of the total fuel $F_{\rm ij}$ and having a constant 
$Q$-factor (eq. 8). Then we apply eq. (7), summing over these sub-phases. 
The SSP total ${\lambda}$-luminosity at the age $i$ is finally
\begin{equation}
L_{\rm i,T}^{\lambda}=\sum_{\rm j}L_{\rm ij}^{\lambda}
\end{equation}
A correct link between MS and PMS luminosity contributions is ensured by 
adopting the same normalization constant $A$ for $\Psi(M)$ in eq. (2) and (3).

The modular structure of the code allows to easily change a particular 
ingredient and/or assumption, independently of the others, as assessing just 
its own impact in the final output (see 5.4 for some examples).

\section{Model Ingredients}

In this section we describe all adopted model ingredients.
The stellar stages included in the synthesis range from the Main 
Sequence to the end of the Thermally Pulsing Asymptotic Giant Branch (TP-AGB).
The Post-AGB has not been taken into account as the $UV$ output of old 
populations is extremely model dependent (cf. Greggio \& Renzini 1990).

\subsection{Stellar input}

For the MS mass-luminosity relation $ L(M,t) $ (eq. 2), a homogeneous set of 
solar chemical composition ($Y=0.27,Z=0.02$) isochrones (Castellani, Chieffi 
\& Straniero 1992, hereafter CCS92), with ages ranging from 30 {\rm Myr} to 10 
{\rm Gyr} has been adopted. An additional isochrone for 15 {\rm Gyr} has 
been constructed from the same set of stellar tracks kindly provided by Oscar
Straniero.
The dwarf-MS component down to 0.1 ${\rm M_{\sun}}$, has been added using the 
models of VandenBerg {\rm et al.} (1983) for $Y$=0.25, $Z$=0.02. 
The small difference in the helium content is not crucial in this context, as 
the location of the lower Main-Sequence, in the $\log L$ - 
$\log T_{\rm e}$ plane, strongly depends on metallicity, but it is rather 
insensitive to the small variations of $Y$ parameter.
\begin{figure} 
\epsffile{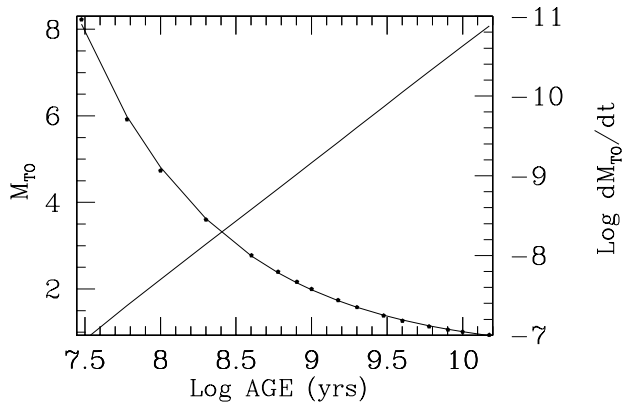}
 \caption{The relation $M_{\rm TO}(t)$ and its time derivative \.M$_{\rm TO}$ 
($t$ in {\rm yr}, $M_{\rm TO}$ in $ {\rm M_{\sun} }$). The relation has been 
obtained via an analytical fit on CCS92 isochrones.
 \label{mtot}}
\end{figure}
 
The relation $ M_{\rm TO}/age $ (Fig. \ref{mtot}) is obtained via an analytical fit 
on CCS92 isochrones 
\begin{equation}
\log M_{\rm TO}(t) = a\log t + b\log^{2} t + c\log^{3} t + d
\end{equation}
where $a$ = 0.212567, $b$ = - 0.108394,  $c$ = 0.005737, $d$ = 2.981730
($M_{\rm TO}$ in ${\rm M_{\sun}}$, $t$ in {\rm yr}). 
The time derivative of $ M_{\rm TO}(t) $ is also displayed in Fig. \ref{mtot}. 
As it is evident, the SSP evolutionary clock, \.M$_{\rm TO}$, is very fast for 
 the large values of $M_{\rm TO}$, then its rate drops greatly as $M_{\rm TO}$
 decreases.
 In fact, $M_{\rm TO}$ ranges from $\simeq$ 8.3 at $t=30 $ {\rm Myr} to 
$\simeq$ 0.94 at $t=15 $ {\rm Gyr} while, in the same age range, \.M$_{\rm TO}$ 
 drops by more than 4 orders of magnitude. This implies a faster SSP luminosity
evolution at young ages ($t\la 1$ {\rm Gyr}).

The stellar evolutionary tracks used to derive eq. (10) have been
constructed using canonical assumptions concerning convective boundaries, i.e.
no overshooting. Slightly higher values of $M_{TO}$ for given age are obtained
 when allowing for convective overshooting, the size of the effect depending
 on a free parameters. However, no general agreement has yet emerged as to what value should
 be assigned to the free parameter, and therefore we prefer to use canonical
 models.
\begin{figure}
 \epsffile{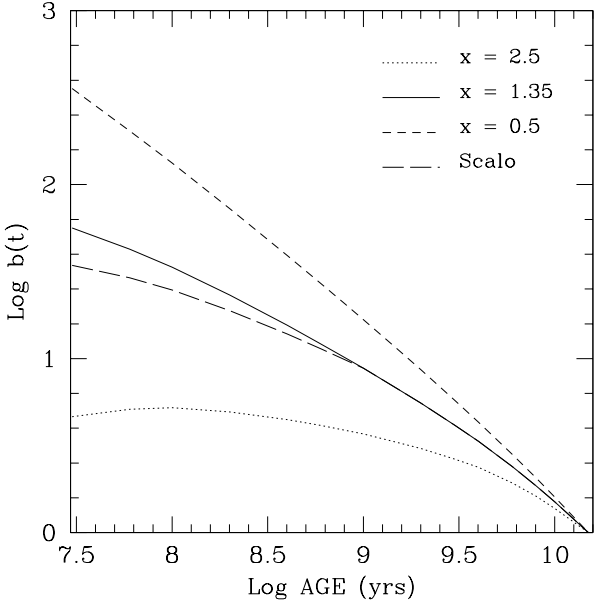}
 \caption{The evolutionary flux $b(t)$ for three choices of the IMF slope $x$.
  The effect of a multislope, like a Scalo type, is also
shown, in which $x$ is -1 for $ M \leq $ 0.3, 1.35 for $ 0.3 < M \leq 2 $,
1.7 otherwise. Functions $b(t)$ are normalized at t = 15 {\rm Gyr}. 
\label{biditi}}
\end{figure}
  
Fig. \ref{biditi} shows the evolutionary flux $ b(t) $, i.e. the rate of 
evolution off the Main Sequence (eq. 4), for four different choices of the
IMF slope. 
The functions $b(t)$ are normalized at $b=1$ for $t=15$ {\rm Gyr}, thus
a comparison between the different slopes is possible regardless of the SSP 
size. 
$b(t)$ is the product of two terms: one increasing ($\Psi$) and one
decreasing (\.M$_{\rm TO}$) with age. The flatter the IMF, the slower the 
increase in time of $\Psi(M)$ and the rapid decrease in \.M$_{\rm TO}$ 
dominates.
This results in a greater SSP luminosity evolution for flatter IMFs.

\subsection{The Fuel Consumption matrix $ F_{\rm ij} $}

Table 1 is our $F_{\rm ij} $ matrix. It gives the amount of hydrogen and helium
 (in $M_{\sun}$) that are burned by a star of mass $M_{\rm {TO},i}$ in each of 
 its PMS stages. The total nuclear fuel follows from eq. (5), while 
$M_{\rm {TO},i}$ and the corresponding age are related by eq. (10).
\begin{table*}
  \centering
 \begin{minipage}{130mm}
  \caption{The $ F_{\rm ij}$  matrix: 
the masses of hydrogen ($m_{\rm H}$) and helium ($m_{\rm He}$) (both in {\rm 
$M_{\odot}$}) burned in each evolutionary PMS $j$ phase, namely SGB, RGB, HB, 
E-AGB, TP-AGB, by a star of $ M=M_{\rm TO,i}$ ($i$ is the age index). Values 
are the same as in Renzini (1992; see the text).} 
   \begin{tabular}{@{}cccccccccc}
& & & & & & & & \\
& & SGB & RGB &\multicolumn{2}{c}{HB}&\multicolumn{2}{c}{E-AGB}&
\multicolumn{2}{c}{TP-AGB} \\ \\
 $AGE$ ({\rm Gyr}) & $M_{\rm TO,i}$ & $m_{\rm H}$ & $m_{\rm H}$ & $m_{\rm H}$ 
 & $m_{\rm He}$ & $m_{\rm H} $ & $m_{\rm He}$ & 
$m_{\rm H}$ & $m_{\rm He}$ \\  
& & & & & & & & \\
0.03 & 8.2247 & 0.0222 &  0.0058 & 0.3116 & 0.8280 & 0.0000 &
0.2091 & 0.0339 & 0.0491 \\
0.06 & 5.9188 & 0.0267 & 0.0059 & 0.1898 & 0.5211 & 0.0000 & 0.4210
& 0.0339 & 0.0491 \\
0.1 & 4.7334 & 0.0296 & 0.0046 & 0.1482 & 0.3993 & 0.0015 & 0.4567
& 0.0339 & 0.0491 \\
0.2 & 3.6001 & 0.0323 & 0.0031 & 0.1233 & 0.3115 & 0.0081 & 0.3918 & 0.0339
& 0.0491 \\
0.4 & 2.7777  & 0.0331 &  0.0041 &  0.1109 &  0.2647 &  0.0110 &
0.3448 & 0.2071 & 0.3138 \\
0.6 & 2.3977 & 0.0329 & 0.0108 &  0.1136 &  0.2366 & 0.0181 & 0.2971 &
0.1931 & 0.2927 \\
0.8 & 2.1640  & 0.0315 & 0.0388 &  0.1108 &  0.2201 &  0.0153 &
0.2867 & 0.1744 & 0.2643 \\
1 & 1.9985  & 0.0307  & 0.0993  & 0.0815 &  0.2349  & 0.0119 &
0.2802 & 0.1427 & 0.2163 \\
1.5 & 1.7408  & 0.0364  & 0.1453 & 0.0668 &  0.2386 &  0.0107 &
0.2969 & 0.0979 & 0.1483 \\
2 & 1.5784 & 0.0431 &  0.1595  & 0.0621  & 0.2373  & 0.0113 &
0.3083 & 0.0747 & 0.1132 \\
3 & 1.3844  & 0.0509 &  0.1689  & 0.0646  & 0.2472  & 0.0139 &
0.2921 & 0.0463  & 0.0701 \\
4 & 1.2651  & 0.0572  & 0.1806 &  0.0618  & 0.2594  & 0.0116  &
0.2827 & 0.0363  & 0.0550 \\
6 & 1.1341  & 0.0625  & 0.1885 &  0.0595 & 0.2665  & 0.0090  &
0.2750 &  0.0254  & 0.0384 \\
8 & 1.0588  & 0.0653  & 0.1921  & 0.0583 &  0.2693  & 0.0076  &
0.2710 & 0.0191 & 0.0289 \\
10 & 1.0110  & 0.0660  & 0.1947  & 0.0560  & 0.2671  & 0.0086 &
0.2671 & 0.0151 &  0.0228 \\
15 & 0.9380  & 0.0659  & 0.1991  & 0.0508  & 0.2598  & 0.0124 &  
0.2598 & 0.0090 & 0.0136 \\
\end{tabular}
\end{minipage}
\end{table*}
The adopted values of $F_{\rm ij}$ in Table 1 are basically the same as those
used by Renzini (1992). They have been derived from
various sets of evolutionary sequences, including Sweigart \& Gross
(1978), Becker (1981), Sweigart, Greggio \& Renzini (1989), Renzini \&
Voli (1981), and Greggio \& Renzini (1990). These data have been
complemented with data from Chieffi \& Straniero (1992, {\it private 
communication}) 
for the fuel consumption during the RGB phase. In principle it would have been
most appropriate to use fuel consumption data from a unique, homogeneous set
of evolutionary calculations. Unfortunately, such a complete set, covering all
evolutionary phases for all stellar masses is still not available. 
However, the present compromise is largely adequate for the scope of
this paper.

A special care was devoted to estimate the fuel consumption during the
TP-AGB phase. Uncertainties in mass loss, mixing, and efficiency of hydrogen
burning at the bottom of the convective envelope (known as envelope burning
process) together prevent pure theory from predicting the amount of
fuel burned during this stage (RB86). The uncalibrated models of
Renzini and Voli (1981, hereafter RV81) used by RB86 largely overestimate the 
TP-AGB contribution for populations younger than $\sim 10^8$ yr in comparison
with the observed contribution in Magellanic Cloud clusters in the
corresponding age range (Frogel, Mould \& Blanco 1990, hereafter FMB90).

The origin of the discrepancy has been controversial for several
years, but a decisive hint came from the discovery that TP-AGB stars
experiencing the envelope burning process do not follow the canonical
AGB core mass-luminosity relation (Bl\"{o}cker \& Sch\"{o}nberner
1991). The 7 $M_{\sun}$ model by Bl\"{o}cker \& Sch\"{o}nberner is much more luminous
and evolves $\sim 10$ times faster than that predicted by the standard core
mass-luminosity relation (e.g. Paczynski 1970a,b) that was used by
RV81. The consequence is a drastic decrease of the
TP-AGB lifetime and of the TP-AGB fuel consumption, that results in a 
lower contribution of this phase to the integrated light of a population.

The transition between TP-AGB stars that experience the envelope
burning process and those which do not is very sharp, and takes place
at a threshold mass that depends sensitively to the mixing length
parameter $\alpha$ (RV81). Now, Magellanic Cloud clusters offer an
opportunity to calibrate this parameter in such a way to reproduce the
observed contribution of AGB stars. Following Bl\"{o}cker \& Sch\"{o}nberner 
(1991) we
assume that TP-AGB stars experiencing envelope burning consume $\sim
10$ times less fuel than predicted by RV81, while we retain RV81 fuel
consumptions for stars that do not experience envelope burning. In
this way the calibration procedure reduces to find the value of $\alpha$ that
best accounts for the data. 
\begin{figure}
 \epsffile{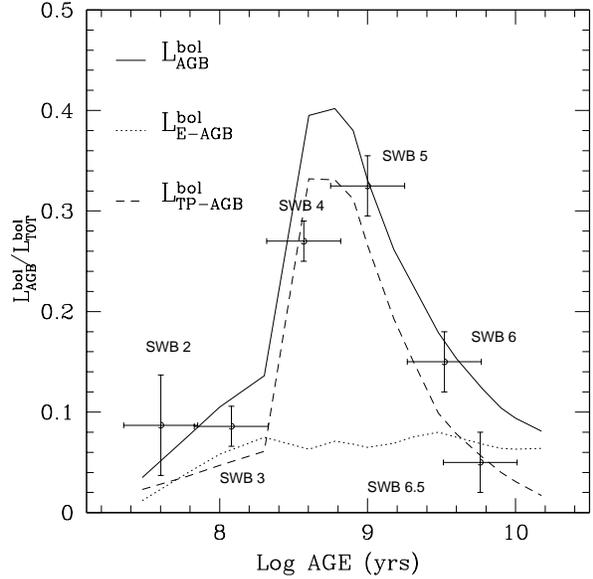}
  \caption{The synthetic AGB bolometric contribution compared with LMC GCs data
  from FMB90. E-AGB and TP-AGB contributions are also shown
  separately. Each data point has been obtained through a procedure aimed at
  minimizing the stochastical fluctuations between clusters having an equal
  SWB parameter (see the text). Error bars indicate the r.m.s.
 \label{agbswb}}
\end{figure}
The result is shown in Fig. \ref{agbswb} which compares
the theoretical AGB contributions for $\alpha=2$ to the Magellanic Cloud 
cluster data from FMB90. For this value of $\alpha$ stars more massive than
$\sim 3$ $ M_{\sun}$ experience the envelope burning process and have the TP-AGB
fuel consumption correspondingly reduced.

In Magellanic Cloud clusters the number and luminosity of AGB stars is
subject to large fluctuations from one cluster to another of similar
age, and therefore so does the AGB contribution to the total
light. This is a result of small number statistics, as clusters contain
at most just a few AGB stars. To cope with this problem, clusters have
been grouped in age bins following the classification of Searle,
Wilkinson and Bagnuolo (1980), with the SWB type being related to age by
the relation given by FMB90. Within each bin the AGB contribution to the 
bolometric luminosity was then obtained by adding together the luminosity of 
all the AGB stars in the binned clusters, and dividing by the sum of all the 
integrated luminosities of the same clusters.
Fig. \ref{agbswb} shows that a jump in the AGB contribution takes place between
SWB type III and SWB type V, or at an age of $\sim 0.2$ Gyr, with the AGB
contribution suddenly increasing from $\sim 10$ per cent to over $\sim 30$ per
cent. This corresponds to the so-called ``AGB phase transition" (RB86) due to 
the development of a bright and well populated TP-AGB phase (see also Renzini
1992).

The TP-AGB phase is populated by {\rm C} and {\rm M} spectral type stars. The 
production of {\rm C} (carbon) stars is a function of both metallicity and the
 initial stellar mass, hence of the SSP age (cf. RV81). 
From FMB90 data for LMC clusters (their Table 4) we have evaluated the 
{\rm C} and {\rm M} star luminosity contributions to the total light for each 
age bin.
As $ L_{\rm TP-AGB}^{\rm bol}\propto F_{\rm TP-AGB} $ (eq. 3), we have 
then calibrated the fractions of TP-AGB fuel burned by {\rm M} and {\rm C} stars
 as functions of SSP age. The use of LMC GCs as calibrators is unavoidable, as 
 they represents the only available sample of clusters covering a broad 
 age-range and being populous enough, so ensuring that the AGB is reasonably 
 well sampled. These clusters have typically $\sim$ 1/2 $Z_{\odot}$ metallicity,
  but a similar sample for solar metallicity is not available. The dependence 
  on $Z$ may be taken into account using the prescriptions of RV81. This will be 
discussed in sec. 5.4.2.
\begin{table}
 \centering
  \caption{\label{tabcmratio}
  The calibration of the TP-AGB fuel (see the text). Column 2 give the 
  total TP-AGB fuel, Column 3 and 4 its fraction burned by {\rm C} and {\rm M} 
  type stars respectively, as functions of the SSP age.}
   \begin{tabular}{@{}cccc}
 & & & \\
 $ AGE (Gyr) $ & $ F_{\rm TP-AGB} $ & $ \% F_{\rm TP-AGB}^{\rm C} $ & 
 $\% F_{\rm TP-AGB}^{\rm M} $ 
 \\
 & & & \\ 
       0.03 &  0.0388 & 0.0000 & 1.0000 \\
       0.06 &  0.0388 & 0.0000 & 1.0000 \\
       0.1 &  0.0388 & 0.0000 & 1.0000 \\
       0.2 &  0.0388 & 0.0000 & 1.0000 \\
       0.4 &  0.2385 & 0.1138 & 0.8862 \\
       0.6 &  0.2224 & 0.1138 & 0.8917 \\
       0.8 &  0.2008 & 0.1160 & 0.9019 \\
       1   &  0.1643 & 0.1421 & 0.8579 \\
       1.5 &  0.1127 & 0.1065 & 0.8935 \\
       2 &  0.0860 & 0.0846 & 0.9154 \\
       3 &  0.0533 & 0.0689 & 0.9311 \\
       4 &  0.0418 & 0.0583 & 0.9417 \\
       6 &  0.0292 & 0.0657 & 0.9343 \\
       8 &  0.0220 & 0.0000 & 1.0000 \\
      10 &  0.0174 & 0.0000 & 1.0000 \\
      15 &  0.0104 & 0.0000 & 1.0000 \\
   \end{tabular}
\end{table}
\begin{figure}
 \epsffile{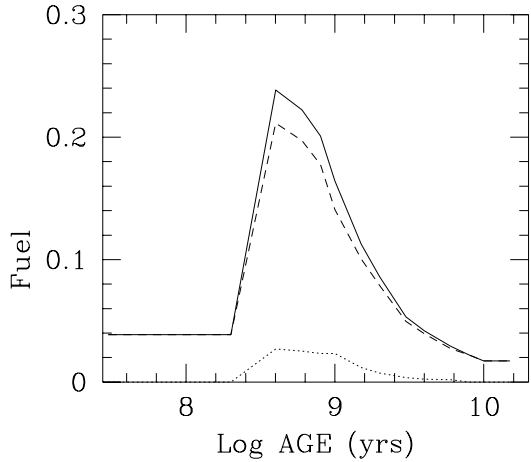}
 \caption{The total amount of TP-AGB fuel ({\it solid line}), 
 its fraction burned by {\rm C} type stars ({\it dotted line }) and by
 {\rm M} type stars ({\it dashed line }), as functions of age (see the
  text).
  \label{cmcal}}
\end{figure}

The calibration is shown in Table 2 (and Fig. \ref{cmcal}). As it can be 
noticed, carbon stars characterize intermediate age SSPs, those having a 
developed TP-AGB. Very old stellar populations (e.g. galactic globular 
clusters and the Galactic Bulge) do not have {\rm C} stars, in agreement with 
theoretical expectations.
\begin{figure}
 \epsffile{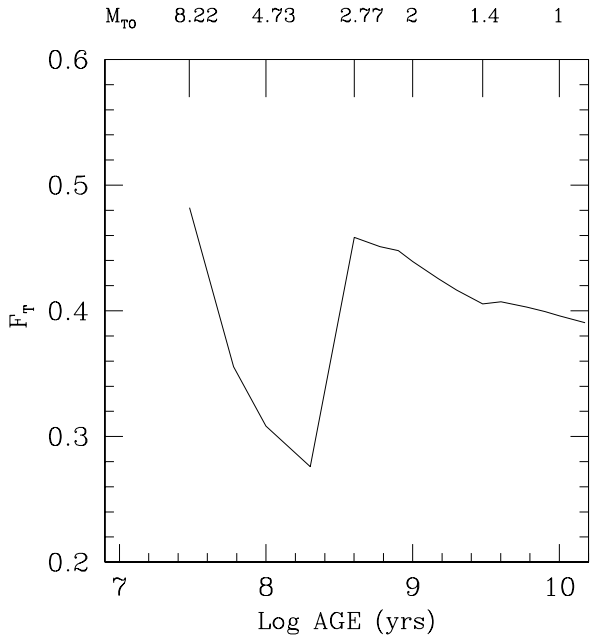}
  \caption{The total fuel consumption (in {\rm $M_{\odot}$}) during the whole 
 post-MS, as a function of the SSP age ({\it lower scale}) 
 and the corresponding turnoff mass ({\it upper scale}).
  \label{ftot}}
\end{figure}

Fig. \ref{ftot} shows the total amount of fuel burned in the whole PMS, 
$ F_{\rm i,T} = \sum_{\rm j} F_{\rm ij}$, as a function of the SSP age and 
the turnoff mass.  
The total fuel decreases with the decreasing mass, till the onset of AGB
phase transition, at ages around 0.2 {\rm Gyr}. Starting from $ t\ga
1$ {\rm Gyr}, $F_{\rm i,T}$ tends to remain nearly constant as the decrease in 
 the TP-AGB fuel is partly compensed by the increase in the fuel burned on the 
 red giant branch (RGB). This epoch is referred to as ``RGB phase transition" (RB86),
 the mass at the transition being around $ \simeq 2.5$ ${\rm M_{\sun}} $, in
 dependence of the chemical composition. 
As it can be noticed, the PMS total fuel depends only weakly on age, 
ranging from $\simeq 0.5$ ${\rm M_{\sun}} $ at $ t \simeq 30
$ {\rm Myr} to $\simeq 0.4$ ${\rm M_{\sun}} $ at $ t \simeq 15 $ {\rm Gyr}. 
Also, it is possible to show that the MS luminosity contribution tends to be 
nearly constant after an age of $ \simeq $ 1 {\rm Gyr}.   
As a consequence, the luminosity evolution of a SSP is mainly
controlled by the evolutionary flux $ b(t) $.  
\begin{figure}
 \epsffile{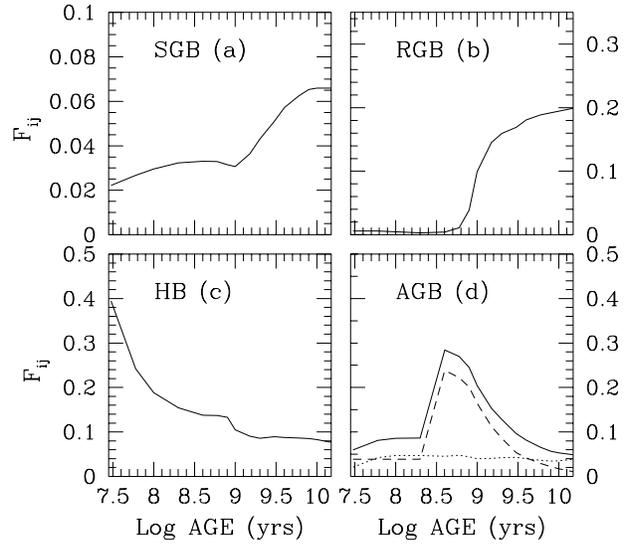}
  \caption{The fuel consumption in each evolutionary PMS phase $j$ as
function of the SSP age; in panel (d), $ F_{\rm E-AGB}$ 
 ({\it dotted line}) and $ F_{\rm TP-AGB}$ ({\it dashed line}) are separately
  shown.
   \label{fgei}}
\end{figure}
Fig. \ref{fgei} shows the time evolution of the fuel consumptions in the
various phases. Worth noting is the sharp increase of the RGB fuel
consumption at $t\simeq 1$ Gyr, which is due to the appearence of
stars developing a degenerate helium core, a phenomenon known as RGB
phase transition (RB86, Sweigart, Greggio \& Renzini 1990). The small
hump in the HB fuel consumption at nearly this same age is also due to
the RGB phase transition, with stars with a minimum value of the core mass
populating the HB, hence experiencing a prolongued HB phase leading to
a local maximum in the hydrogen consumption (cf. Table 1).
\begin{figure}
 \epsffile{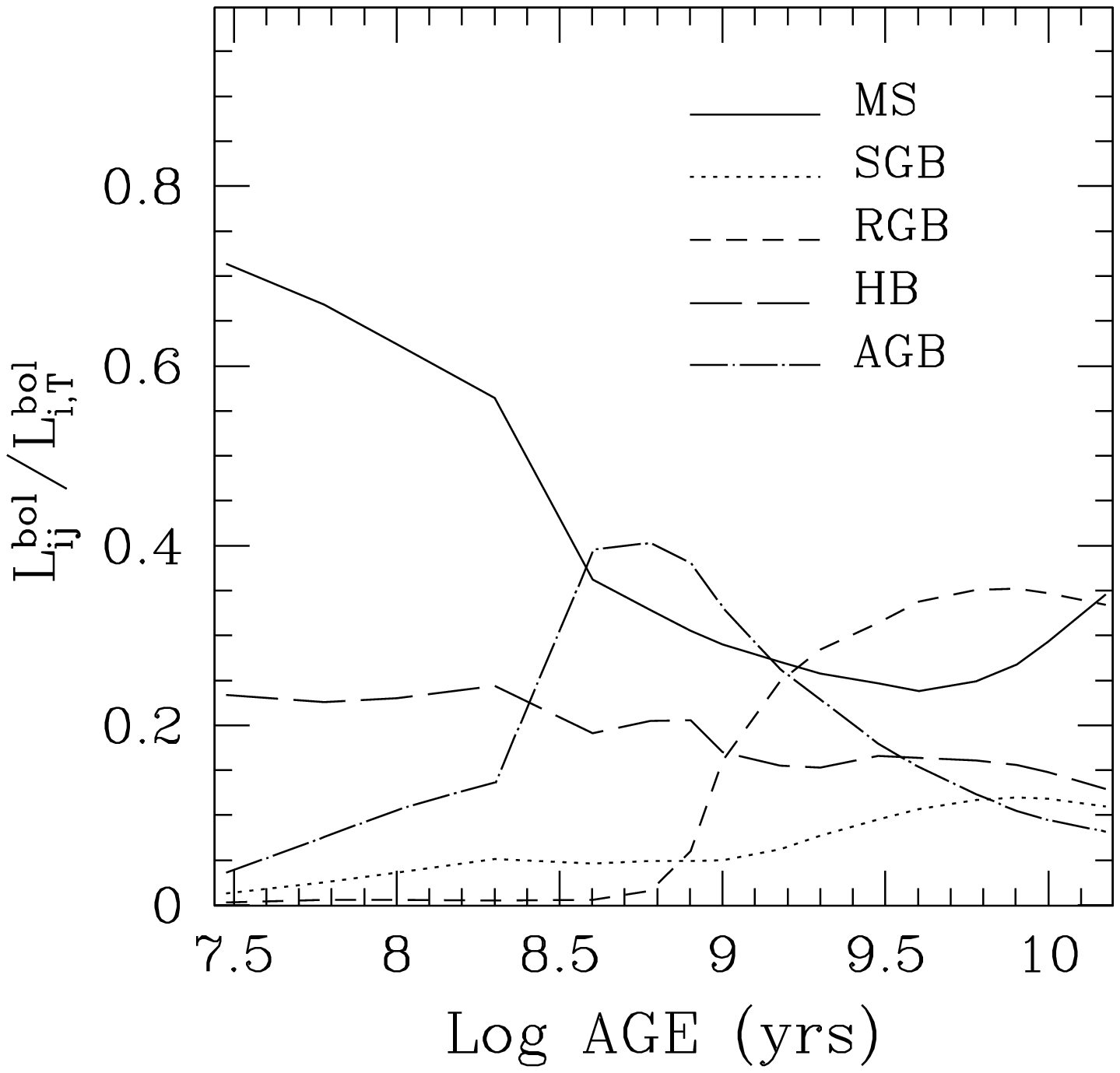}
  \caption{The time evolution of the relative contributions
 $L_{\rm ij}^{bol}/L_{\rm i,T}^{bol}$ of stars in the various
  evolutionary stages to the integrated bolometric light of a stellar 
  population.
\label{contrbol}}
\end{figure}

Finally, Fig. \ref{contrbol}  shows the contribution of various evolutionary
stages to the total bolometric luminosity, $L_{\rm ij}^{\rm bol}/L_{\rm
i,T}^{\rm bol}$, obtained with the procedure described in sec. 2 and the
ingredients of sec. 3.1 and 3.2. The energetic of young SSPs ($t\la$ 0.3 
{\rm Gyr}) is dominated by MS stars, that of intermediate age ones 
(0.3 $\la t\la$ 3 {\rm Gyr}) by AGB stars and, finally, by RGB stars for $t\ga$
 3 {\rm Gyr}. For ages older than $\simeq $ 6 {\rm Gyr}, the relative MS 
 contribution tends to rise. This effect is caused by the MS integrated 
 luminosity decreasing slower than $ b(t) $. As the total fuel keeps nearly 
 constant, the net result is a lower total luminosity as mainly due to a lower 
 PMS luminosity.
\begin{table*}
 \centering
  \begin{minipage}{120mm}
   \caption{The $ {(T_{\rm e},g)}_{\rm ij}$ matrix for the 30 {\rm Myr}, the 1
   {\rm Gyr} and the 15 {\rm Gyr} SSP: effective temperatures and surface gravities
for the $j$ evolutionary phases, as functions of the $i$ ages. 
Values have been drawn from CCS92 isochrones.}   
    \begin{tabular}{@{}cccccccccc}
 \multicolumn{2}{c}{MS} &\multicolumn{2}{c}{SGB}
 &\multicolumn{2}{c}{RGB} & \multicolumn{2}{c}{HB} & 
\multicolumn{2}{c}{E-AGB} \\
 & & & & & & & & &  \\
$\log g$ & $\log T_{\rm e}$ & $\log g$ &  $\log T_{\rm e}$ & $\log g$ & 
$ \log T_{\rm e}$ & $\log g$ & $\log T_{\rm e}$ & $\log g$ & $\log T_{\rm e}$ 
 \\  
 & & & & & & & & &  \\
\multicolumn{10}{c}{AGE = 30 {\rm Myr}} \\
 & & & & & & & & &  \\
4.76 & 3.6044 & 3.77 & 4.3148 & 1.42 & 3.6537 & 0.88 & 3.6092 &
  0.89 & 3.6154 \\

4.59 & 3.7128 & 3.76 & 4.3120 & 1.32 & 3.6397 & 0.95 & 3.6206 &
  0.70 & 3.5998  \\
4.38 & 3.8055 & 3.75 & 4.3095 & 1.14 & 3.6248 & 0.96 & 3.6235 &
  0.75 & 3.6032 \\
4.32 & 3.8880 & 3.74 & 4.3075 & 0.99 & 3.6144 & 1.03 & 3.6554 &
  0.65 & 3.5956 \\ 
4.33 & 3.9614 & 3.73 & 4.3059 & 0.93 & 3.6100 & 1.64 & 3.8188 &
  0.56 & 3.5890 \\
4.33 & 4.0165 & 3.63 & 4.2824 & 0.90 & 3.6082 & 1.75 & 3.8491 &
  0.48 & 3.5838 \\
4.31 & 4.0585 & 3.53 & 4.2636 & 0.89 & 3.6071 & 1.83 & 3.8723 &
  0.42 & 3.5794 \\
4.29 & 4.0949 & 3.45 & 4.2457 & 0.87 & 3.6064 & 1.90 & 3.8925 &
  0.35 & 3.5749 \\
4.27 & 4.1269 & 3.37 & 4.2259 & 0.87 & 3.6058 & 1.96 & 3.9114 &
  0.28 & 3.5704  \\ 
4.25 & 4.1545 & 3.27 & 4.2020 & 0.86 & 3.6052 & 1.98 & 3.9191 &
  0.20 & 3.5650 \\
4.23 & 4.1781 & 3.15 & 4.1709 & 0.85 & 3.6046 & 2.02 & 3.9315 & &
  \\ 
4.20 & 4.1991 & 2.99 & 4.1274 & 0.84 & 3.6040 & 2.07 & 3.9473 & &
  \\
4.17 & 4.2178 & 2.76 & 4.0594 & 0.84 & 3.6036 & 2.06 & 3.9451 & &
  \\
4.14 & 4.2338 & 2.33 & 3.9343 & 0.83 & 3.6033 & 2.05 & 3.9419  & & \\ 
4.10 & 4.2480 & & & 0.83 & 3.6031 & 2.04 & 3.9383  & & \\
4.06 & 4.2602 & & & 0.83 & 3.6029 & 2.02 & 3.9343  & &  \\
4.01 & 4.2701 & & & 0.82 & 3.6028 & 1.99 & 3.9273 & & \\
3.95 & 4.2771 & & & 0.82 & 3.6027 & 1.94 & 3.9156  & & \\
3.88 & 4.2800 & & & 0.82 & 3.6026 & 1.88 & 3.8980 & &  \\
3.78 & 4.2776 & & & 0.82 & 3.6025 & 1.77 & 3.8704 & & \\ 
3.67 & 4.2725 & & & 0.82 & 3.6024 & 1.60 & 3.8225  & & \\
3.67 & 4.2767 & & & 0.82 & 3.6024 & 1.12 & 3.6905 & & \\
3.77 & 4.3161 & & & 0.82 & 3.6023 & & & &  \\ 
 & & & & & & & & &  \\
\multicolumn{10}{c}{AGE = 1 {\rm Gyr}} \\
 & & & & & & & & &  \\
4.76 & 3.6044 & 3.85 & 1.4166 & 3.42 & 3,7177 & 2.71 & 3.6860 & 2.45 &
  3.6764 \\
4.71 & 3.6194 & 3.83 & 3.8927 & 3.34 & 3.7086 & 2.71 & 3.6876 & 2.10 &
  3.6562 \\
4.66 & 3.6416 & 3.82 & 3.8901 & 3.25 & 3.7031 & 2.71 & 3.6881 & 2.19 &
  3.6617 \\
4.62 & 3.6706 & 3.80 & 3.8872 & 3.16 & 3.6990 & 2.72 & 3.6885 & 2.19 &
  3.6618 \\
4.60 & 3.6990 & 3.78 & 3.8842 & 3.08 & 3.6957 & 2.72 & 3.6888 & 2.18 &
  3.6614 \\
4.56 & 3.7240 & 3.76 & 3.8801 & 3.00 & 3.6925 & 2.72 & 3.6891 & 2.16 &
  3.6604 \\
4.52 & 3.7455 & 3.74 & 3.8759 & 2.93 & 3.6897 & 2.72 & 3.6894 & 2.13 &
  3.6584 \\
4.47 & 3.7618 & 3.71 & 3.8692 & 2.86 & 3.6867 & 2.72 & 3.6897 & 2.08 &
  3.6553 \\
4.42 & 3.7770 & 3.67 & 3.8604 & 2.79 & 3.6842 & 2.72 & 3.6899 & 1.98 &
  3.6498 \\
4.37 & 3.7914 & 3.63 & 3.8475 & 2.73 & 3.6813 & 2.72 & 3.6901 & 1.81 &
  3.6401 \\
4.34 & 3.8066 & 3.56 & 3.8284 & 2.67 & 3.6788 & 2.72 & 3.6903 & 1.32 &
  3.6099 \\
4.30 & 3.8201 & 3.48 & 3.8029 & 2.60 & 3.6758 & 2.71 & 3.6905 \\
4.26 & 3.8318 & 3.42 & 3.7735 & 2.55 & 3.6758 & 2.71 & 3.6908 \\
4.22 & 3.8415 & 3.42 & 3.7425 & 2.49 & 3.6708 & 2.70 & 3.6910 \\
4.19 & 3.8525 & & & 2.44 & 3.6708 & 2.69 & 3.6909 \\ 
4.16 & 3.8619 & & & 2.40 & 3.6666 & 2.68 & 3.6907 \\  
4.12 & 3.8687 & & & 2.35 & 3.6641 & 2.67 & 3.6901 \\
4.08 & 3.8727 & & & 2.31 & 3.6626 & 2.65 & 3.6893 \\
4.01 & 3.8726 & & & 2.11 & 3.6516 & 2.62 & 3.6872 \\
3.93 & 3.8666 & & & 1.98 & 3.6451 \\
3.83 & 3.8571 & & & 1.64 & 3.6260 \\
3.83 & 3.8607 & & & 1.37 & 3.6103 \\
3.91 & 3.9057 & & & 0.97 & 3.5858 \\
  \end{tabular}
 \end{minipage}
\end{table*}
\begin{table*}
   \contcaption {} 
    \begin{tabular}{@{}cccccccccc}
 \multicolumn{2}{c}{MS} &\multicolumn{2}{c}{SGB}
 &\multicolumn{2}{c}{RGB} & \multicolumn{2}{c}{HB} & 
\multicolumn{2}{c}{E-AGB} \\
 & & & & & & & & &  \\
$\log g$ & $\log T_{\rm e}$ & $\log g$ &  $\log T_{\rm e}$ & $\log g$ & 
$ \log T_{\rm e}$ & $\log g$ & $\log T_{\rm e}$ & $\log g$ & $\log T_{\rm e}$  \\  
 & & & & & & & & &  \\
\multicolumn{10}{c}{AGE = 15 {\rm Gyr}} \\
 & & & & & & & & &  \\
4.74 & 3.6065 & 4.29 & 3.7455 & 3.93 & 3.7102 & 2.26 & 3.6607
  & 2.02 & 3.6486  \\
4.70 & 3.6175 & 4.23 & 3.7458 & 3.84 & 3.6961 & 2.29 & 3.6621
  & 1.53 & 3.6183 \\ 
4.66 & 3.6298 & 4.24 & 3.7461 & 3.81 & 3.6938 & 2.28 & 3.6630
  & 1.69 & 3.6294 \\  
4.68 & 3.6539 & 4.22 & 3.7464 & 3.79 & 3.6920 & 2.30 & 3.6640
  & 1.71 & 3.6305 \\
4.66 & 3.6632 & 4.18 & 3.7456 & 3.76 & 3.6901 & 2.31 & 3.6649
  & 1.71 & 3.6302 \\  
4.65 & 3.6681 & 4.17 & 3.7445 & 3.73 & 3.6888 & 2.31 & 3.6657
  & 1.70 & 3.6298\\  
4.64 & 3.6732 & 4.15 & 3.7429 & 3.69 & 3.6876 & 2.31 & 3.6665
  & 1.68 & 3.6283 \\ 
4.63 & 3.6783 & 4.13 & 3.7407 & 3.66 & 3.6864 & 2.34 & 3.6670
  & 1.62 & 3.6261 \\      
4.62 & 3.6835 & 4.09 & 3.7379 & 3.63 & 3.6858 & 2.32 & 3.6674
  & 1.60 & 3.6231 \\   
4.61 & 3.6887 & 4.07 & 3.7344 & 3.59 & 3.6850 & 2.32 & 3.6679 & 
 1.52 & 3.6191 \\    
4.59 & 3.6993 & 4.02 & 3.7298 & 3.55 & 3.6843 & 2.31 & 3.6680
 & 1.46 & 3.6142 \\     
4.58 & 3.7040 & 3.99 & 3.7223 & 3.51 & 3.6836 &2.31& 3.6677 \\        
4.57 & 3.7091 & & & 3.47 & 3.6829 &2.31 & 3.6676 \\        
4.56 & 3.7136 & & & 3.42 & 3.6819 &2.30 & 3.6674\\        
4.54 & 3.7179 & & & 3.37 & 3.6811 &2.29 & 3.6673\\        
4.53 & 3.7222 & & & 3.32 & 3.6802 &2.28&3.6668\\        
4.51 & 3.7263 & & & 3.26 & 3.6794 &2.27&3.6661 \\       
4.49 & 3.7303 & & & 3.17 & 3.6774 &2.26&3.6652\\       
4.47 & 3.7340 & & & 3.12 & 3.6761 &2.23&3.6636\\         
4.45 & 3.7375 & & & 3.03 & 3.6742 &2.18&3.6596\\        
4.43 & 3.7405 & & & 2.90 & 3.6712 \\        
4.40 & 3.7436 & & & 2.77 & 3.6674 \\        
4.39 & 3.7445 & & & 2.63 & 3.6616 \\        
4.38 & 3.7450 & & & 2.56 & 3.6597 \\        
4.35 & 3.7469 & & & 2.32 & 3.6504 \\        
4.32 & 3.7479 & & & 1.86 & 3.6283 \\        
4.27 & 3.7484 & & & 0.02 & 3.5179 \\       
  \end{tabular}
\end{table*}

\subsection{The temperature/gravity matrix $ (T_{\rm e},g)_{\rm ij}$}

The $ (T_{\rm e},g)_{\rm ij}$ matrix contains the distribution of effective 
temperatures and surface gravities of the evolutionary phases $j$ at
the SSP age $i$.

Stellar effective temperatures depend crucially on the efficiency of convective
 energy transfer, parametrized by the mixing lenght parameter ${\alpha}$. 
 The use of uncalibrated theoretical effective temperatures in EPS
 computations is extremely dangerous, as the calibration of ${\alpha}$ is the
 synchronization of the EPS clock. CCS92 isochrones are computed 
 for ${\alpha}=1.6$, a value that matches the observed Sun. Moreover, these 
 isochrones have been tested on a sample of well-studied Galactic clusters and  
 account well for the observed location of stars in the various evolutionary stages 
 (see CCS92). The $ (T_{\rm e},g)_{\rm ij}$ values for MS to E-AGB have been 
 taken from CCS92 calibrated isochrones. V83 models have been used for the 
 dwarf-Main Sequence component ($0.1 \leq M < 0.6$).

For the TP-AGB phase, the relative proportions of {\rm C} and {\rm M} 
stars as functions of age (from FMB90, see sec. 3.2) are used as indicators 
of the evolution of the spectral type along the TP-AGB. 
\begin{figure}
 \epsffile{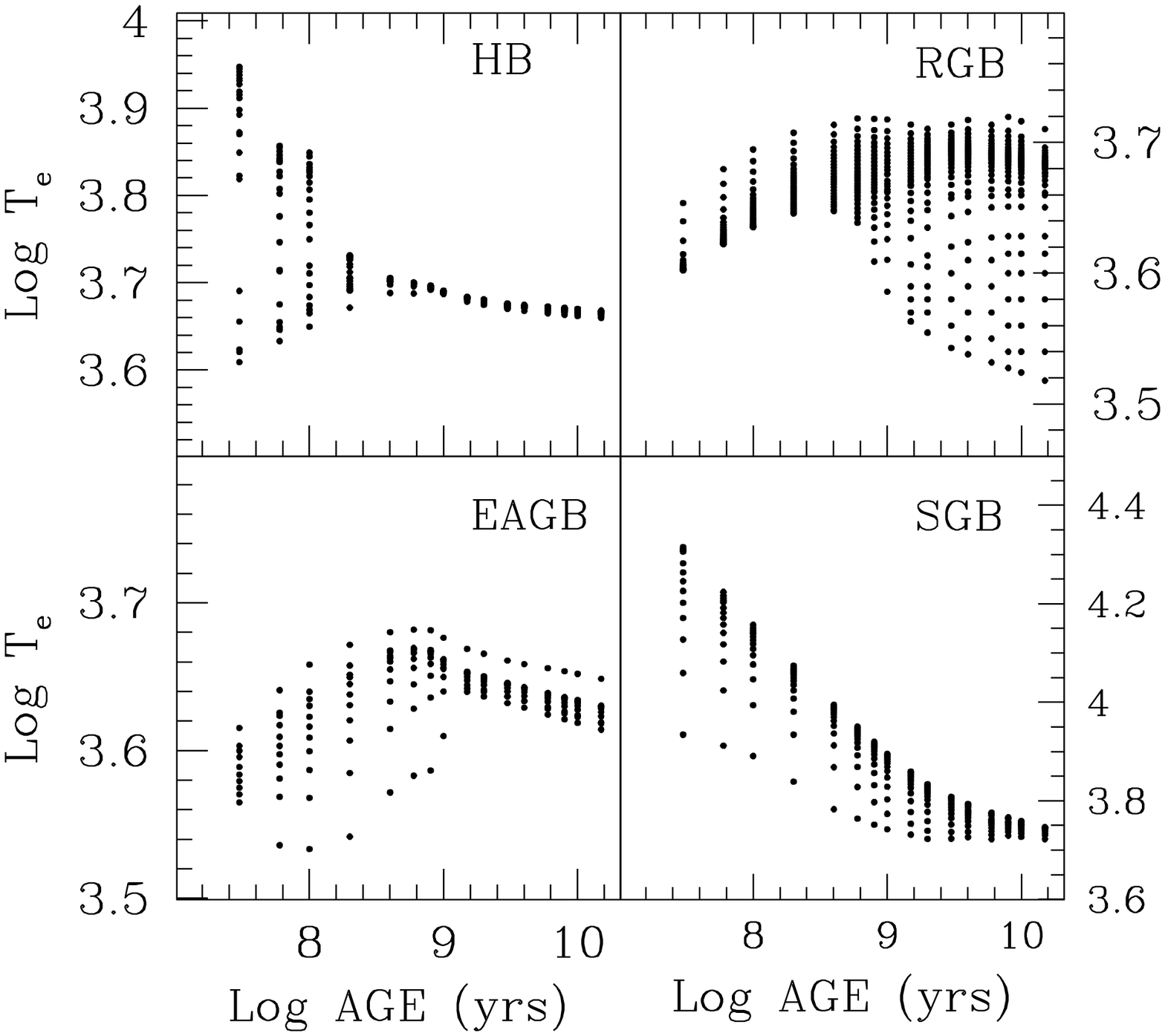}
 \caption{The time evolution of effective temperatures $T_{\rm e}$ in the 
 evolutionary PMS stages.
 \label{tepms}}
\end{figure}
\begin{figure}
 \epsffile{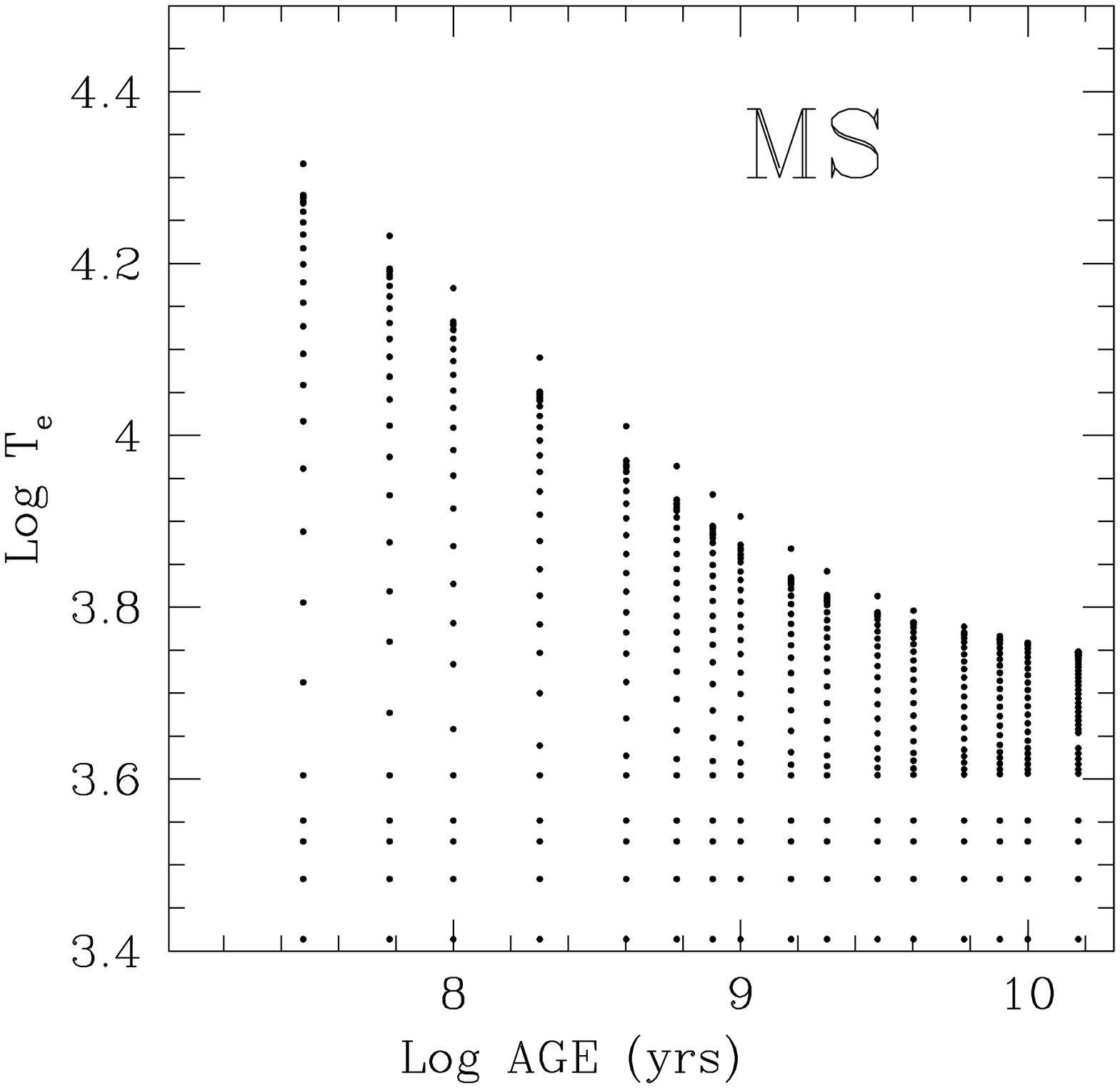}
 \caption{The same as Fig. 8 for the Main Sequence.
 \label{tems}}
\end{figure}

The time evolution of the $T_{\rm e}$ ranges spanned by PMS evolutionary 
stages is displayed in Fig. \ref{tepms}, while Fig. \ref{tems} shows the same 
for the MS. To note in Fig. \ref{tepms} the red HB clump at late ages 
(see 5.4.1). As an example, Table 3 is the $ (T_{\rm e},g)_{\rm ij}$ matrix for 
$i$=30 {\rm Myr}, 1 {\rm Gyr} and 15 {\rm Gyr} ($M_{\rm TO}$ = 8.2247, 1.9985
and 0.9380 respectively). 
By changing the content of the $ (T_{\rm e},g)$ matrix, it is possible to
infer the impact on the integrated photometric features of SSPs, of some model
parameters such as mixing length and mass loss, that directly influence 
effective temperatures and surface gravities (see 5.4).  

\subsection{The colour/bolometric correction matrix}   

The $ (CI,BC)_{\rm lm} $ matrix contains bolometric corrections and colour 
indices, to distribute the total energy through the various 
passbands. It covers the range of $ (T_{e},g)_{\rm ij}$ matrix and 
contains empirical colours suitable to TP-AGB {\rm M} and {\rm C} type stars.
\begin{table}
 \centering
  \caption{ The adopted solar normalization}
   \begin{tabular}{@{}ccccccc}
 $BC_{V\odot} $ & $ M_{bol\odot}$ & $ M_{U\odot} $ & $ M_{B\odot} $ & $
  M_{V\odot}$ & $ M_{R\odot}$ & $ M_{K\odot}$ \\  
 -0.08 & 4.75  &  5.61 & 5.48 & 4.83 & 4.31 & 3.41 \\
\end{tabular}
\end{table}
$BC_{V}$ and $(U-B)$ are taken from Kurucz (1992) theoretical
grids, on which a quadratical interpolation in $\log T_{\rm e}$, $\log g$ has 
been performed. 
For $ T_{\rm e} \leq 3500 $ {\degr {\rm K}, the limit of Kurucz models, 
some extrapolations have been necessary. 
$(B-V)$ colours have been taken from CCS92 isochrones. They are computed on 
the basis of the empirical color-temperature relation obtained by 
Arribas \& Martinez (1988, 1989), for 4000 $K$ $\leq T_e \leq$ 8000 $K$ (see 
the authors for more details); outside this $T_{\rm e}$ range, they are based 
on Kurucz models.
$(V-R)$ and $(V-K)$ colours, for spectral types earlier than {\rm K}, have 
been taken from Johnson (1966) empirical tables, on which a linear 
interpolation in $(B-V)$ has been performed.
For {\rm M} type stars, the theoretical models by Bessel {\rm et al.} (1989) 
have been used. Colours have been linearly interpolated in $T_{\rm e}$.
For {\rm M} and {\rm C} type TP-AGB stars $(V-R)$ and $(V-K)$ values have been 
taken from various sets of observational data (Cohen {\rm et al.} 1981; Cohen 
1982; Westerlund {\rm et al.} 1991).
Finally, Table 4 shows the adopted values for the solar normalization (from
Allen 1991). By changing the $ (CI,BC)_{\rm lm} $ matrix, it is possible to 
explore the influence on model output of the photometric ingredients and the associated 
uncertainties. 

\section{Results}

Evolutionary synthesis models for SSPs have been computed with the algorithm 
presented in Section 2 and the ingredients described in Section 3. 
Models are characterized by a solar chemical composition and ages ranging from 
30 {\rm Myr} to 15 {\rm Gyr}. 
The model predicts the SSP photometric properties as 
functions of age, i.e. the broad-band colours $(U-V)$, $(U-B)$, $(B-V)$, 
$(V-R)$ and $(V-K)$, the bolometric corrections, the specific contributions of 
evolutionary phases to the bolometric and to the broad band luminosities, and 
the stellar mass-to-light ratios (see App. A, Tables A1, A2 for a complete 
list). 

\subsection{The time evolution of integrated colours}

\begin{figure}
 \epsffile{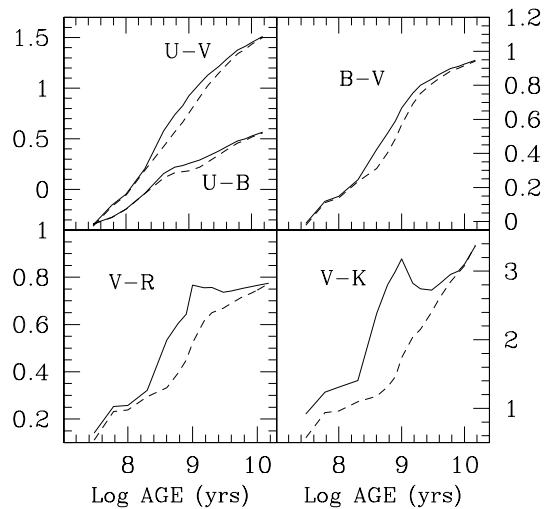}
  \caption{The time evolution of synthetic broad-band colours for a Salpeter 
  IMF. Dashed lines represent models in which the TP-AGB contribution is not
 taken into account.
\label{colorage}}
\end{figure}
The time evolution of synthetic broad-band colours is shown in Fig.
 \ref{colorage}. 

$(U-V)$, $(U-B)$ and $(B-V)$ colours vary smoothly with age, because they 
trace the turnoff effective temperature.
The evolutionary trend of IR colours is instead controlled by the AGB and RGB 
phase transitions. Their impact is particularly evident in $(V-K)$, rising
from $\sim 1.3$ to $\sim 3.2$ in a short time interval. Its early rise is due 
to the AGB phase transition starting at $ t\simeq 250 $ {\rm Myr}. The RGB 
phase transition takes place at $t\simeq$  500 {\rm Myr}, contributing to keep 
values high of $(V-K)$. The modest decrease in $(V-K)$ at 1 {\rm Gyr} $ \la t \la 
4 $ {\rm Gyr} reflects the reduction of the TP-AGB contribution.
The TP-AGB phase plays a major role in determining the IR
colours at intermediate ages. This can be appreciated looking at models in
which the TP-AGB contribution has been omitted (also shown in Fig.
\ref{colorage}).
The impact of the two phase transitions on optical colours is much weaker as 
AGB and RGB stars radiate mostly in the near infrared.
\begin{figure*}
 \epsffile{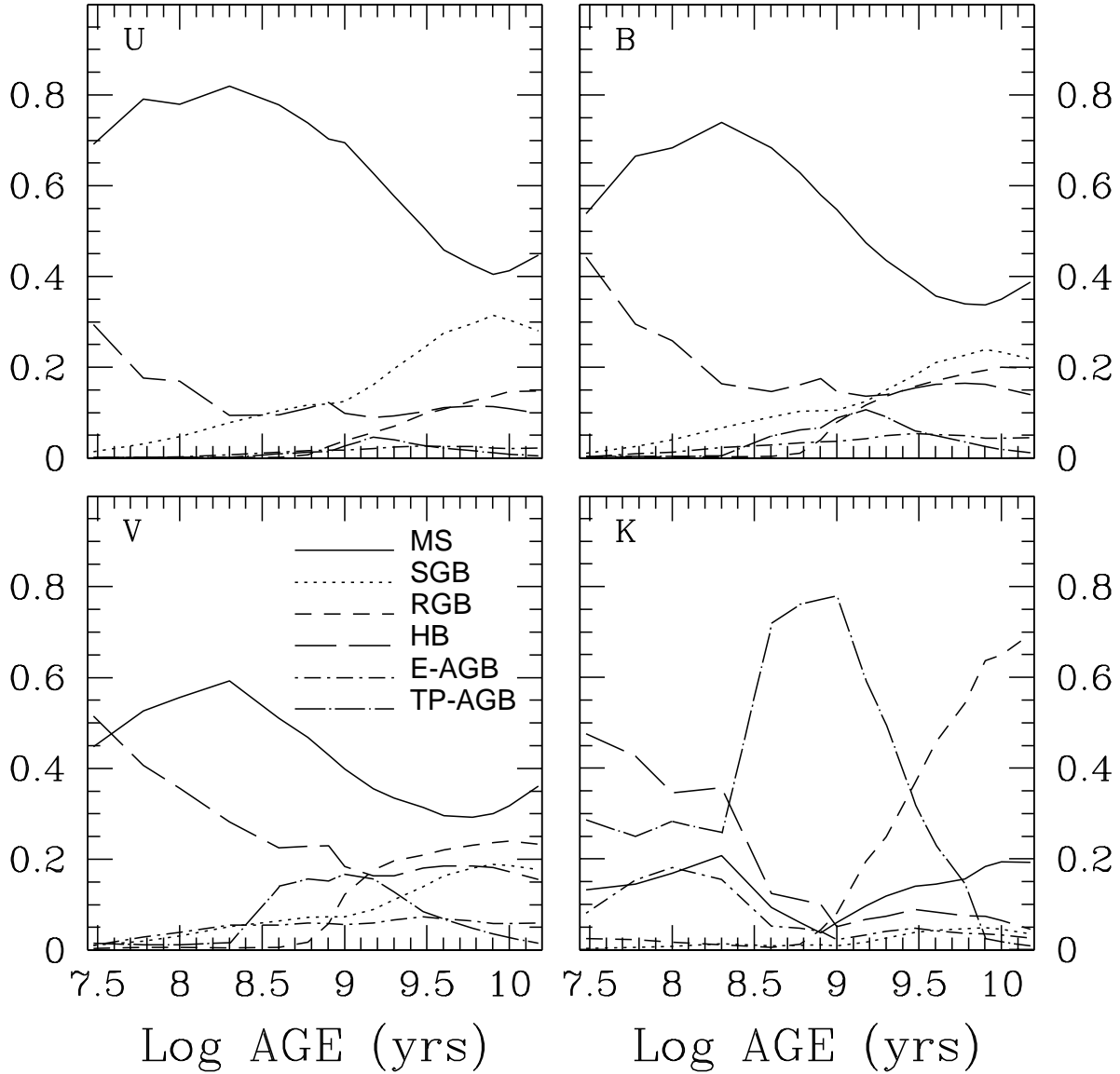}
  \caption{The evolutionary phase contributions to the total 
 monochromatic luminosities, $ L_{\rm ij}^{\lambda}/L_{\rm i,T}^{\lambda} $,
 as functions of the SSP age, for a Salpeter IMF.
  \label{contrmono}}   
\end{figure*}

The evolutionary behaviour of integrated colours can be easily understood 
by looking at the luminosity contributions of evolutionary 
stages, $ L_{\rm ij}^{\lambda}/L_{\rm i,T}^{\lambda} $, shown in Fig.
\ref{contrmono}. 
The UBV emission mainly comes from MS stars at any age, except for 
$t \la 60 $ {\rm Myr}, when the major $V$ contributors are {\rm He}-burning 
supergiants of intermediate spectral type.
The $K$ band reveals the AGB phase transition at intermediate ages, when
the TP-AGB contribution rises from $\simeq 20$ per cent to $\simeq 80 $ per
cent. The RGB phase transition is visible in all bands, but particularly 
in $K$, where the RGB contribution dominates for t $ \geq$ 3 {\rm Gyr}. 
\begin{figure}
 \epsffile{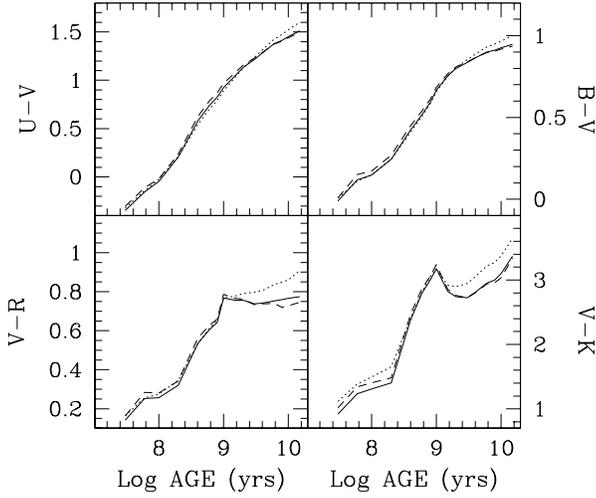}
 \caption{The influence of IMF on integrated broad-band
  colours: a Salpeter ({\it solid line}), a dwarf-dominated
  ({\it dotted line}) and a giant-dominated ({\it dashed line}) IMF.
    \label{colorimf}}
\end{figure}
\begin{figure}	
 \epsffile{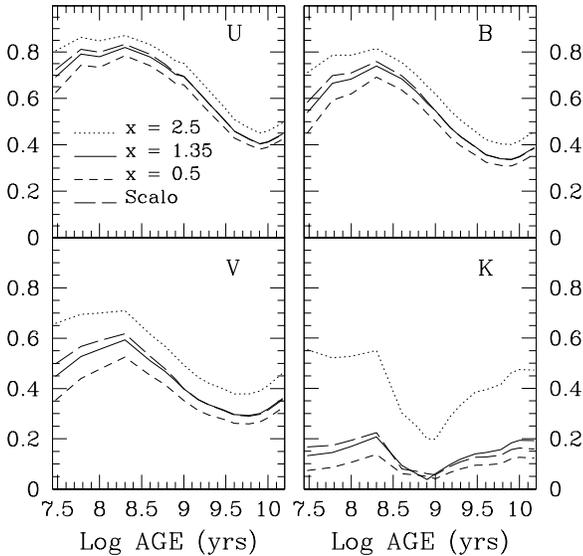}
  \caption{The dependence of MS luminosity contributions on the IMF slope, 
from a dwarf-dominated ($x$=2.5) to a giant-dominated ($x$=0.5) one; $x$=1.35 
is the classical Salpeter and Scalo refers to a three-slope IMF, defined in 
the caption of Fig. 2.
  \label{contrmsimf}}
\end{figure}
The IMF has no influence on integrated optical colours (Fig. \ref{colorimf}),
 as the variations induced in $U,B$ and $V$ luminosities are nearly 
 identical, leaving the corresponding colours unchanged.
A small dependence is visible only in
the infrared, at ages greater than $\simeq 1 $ {\rm Gyr}, when a 
dwarf-dominated ($x=2.5$) SSP has colours redder than a Salpeter one, because
of the contribution of low-mass stars.   
The Main Sequence contribution is the most sensitive to the IMF slope (see 
Fig. \ref{contrmsimf}).

\subsection{The IMF scale factor}

The size of a stellar population is direclty connected to the scale factor 
$A$ of the IMF. Thus it is useful to express $A$ in terms of the integrated 
luminosity, which is also proportional to the size of a SSP and directly 
observable. To this aim, we solve eq. (4) for $A$, obtaining 
\begin{equation}
 A = B(t)L_{\rm T}^{\rm bol}(t){M_{\rm TO}}^{1+x}{|{{dM}{_{\rm TO}}\over{dt}}|}^{-1}  
\end{equation}
where $ B(t)= b(t)/L_{\rm T}^{\rm bol}(t) $ is the evolutionary flux per unit 
luminosity of the SSP (the `{\it specific evolutionary flux}'), a quantity 
nicely independent of the IMF slope (RB86).
The total bolometric luminosity $ L_{\rm T}^{\rm bol}(t) $ is related to the 
total monochromatic ones $ L_{\rm T}^{\lambda} $(t), through the bolometric 
correction factors $ B_{\rm c}^{\lambda} $(t)
\begin{equation}
L_{\rm T}^{\rm bol}(t) = B_{\rm c}^{\lambda}(t) L_{\rm T}^{\lambda}(t).   
\end{equation}
$ B_{\rm c}^{\lambda} $ are functions of metallicity, IMF slope and age and are
the direct product of synthetic SSP models. So, through eq. (11) and
 (12), we get $A$ as a function of the luminosity in a certain band. 
\begin{figure}
 \epsffile{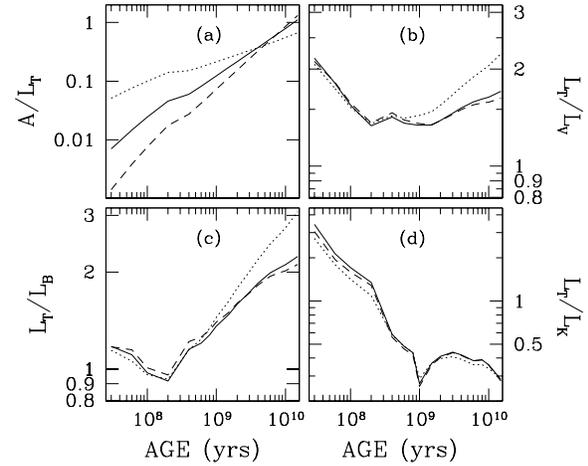}
  \caption{The relation between the scale factor $A$ and the bolometric 
luminosity of a SSP, $A/L_{\rm T}^{\rm bol}$ (panel a). Also shown are the 
bolometric correction  factors $ B_{\rm c}^{\rm {\lambda}} $ (panels b,c,d),  
for three different choices of the IMF slope. Solid line refers to a Salpeter,
dotted line to a dwarf-dominated ($x$ = 2.5) and dashed line to
a giant-dominated ($x$ = 0.5) IMF.
   \label{scalefactor}}
\end{figure}

The ratio between the scale factor $A$ and the bolometric luminosity, and
the bolometric correction factors $ B_{\rm c}^{\lambda} $ are shown in Fig. 
\ref{scalefactor}, all as functions of age and IMF slope. 
The $ L_{\rm T}^{\rm bol}/L_{\rm T}^{\rm {\lambda}}$ ratios are fairly 
independent of the IMF slope, as far as the IMF is not too steep for masses 
below $\simeq $ 0.5 ${\rm M_{\odot}} $ (i.e., $ 1+x \leq 2.5 $). 
This is a consequence of the very narrow mass range producing most of the 
luminosity at nearly any SSP ages (see sec. 2). As an example, by adopting a
Salpeter IMF, we obtain $ A\simeq 1.12\cdot L_{\rm T}^{\rm bol}\simeq 2.5\cdot 
L_{\rm T}^{\rm B}$ for a 15 {\rm Gyr} SSP. 

\subsection{The stellar mass-to-light ratio}

At any age, an SSP is composed of `living' stars and `dead-remnants', 
according to the initial mass of stars.
The stellar mass is given by the convolution of the present mass $M(t)$ with the IMF. The present mass $M(t)$ coincides with 
the initial mass ($M(t)$ = $M_{\rm in}$) for $M_{\rm in} \leq M_{\rm TO} (t)$ 
while $M(t)$ = $M_{\rm R}$ (the remnant mass) for $M_{\rm in} > 
M_{\rm TO} (t)$. The kind of remnant also depends on the initial stellar mass:
 here we adopt the following recipe (cf. Renzini \& Ciotti 1993).
For $ M_{\rm TO}(t) \leq M_{\rm in} < 8.5 $, the remnant is a white dwarf of 
mass $M_{\rm R}$ = 0.077 $M_{\rm in}$+ 0.48; for $ 8.5 \leq M_{\rm in} < 40 $, stars 
leave a 1.4 $ {\rm M_{\odot}} $ neutron star; finally, 
stars with $ M_{\rm in} \geq 40 $ are assumed to leave massive black-holes 
of mass $M_{\rm R}$ = 0.5 $ M_{\rm in} $.
\begin{figure}
 \epsffile{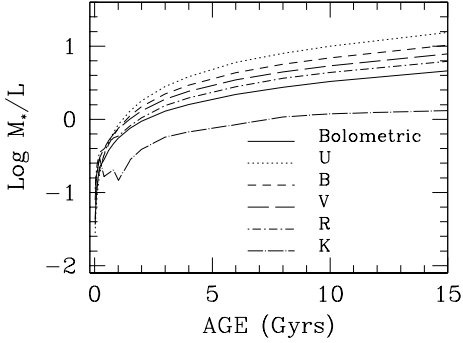}
  \caption{The evolution of stellar mass-to-light ratios for a Salpeter IMF.
   \label{ml}}
\end{figure}

Fig. \ref{ml} shows the evolution of stellar mass-to-light ratios for the
bolometric and $U,B,V,R$ and $K$ luminosities. The early time evolution of $ M_{*}/L$ is very fast,
 then, after an age of $\simeq 5 $ {\rm Gyr}, it becomes mild. For example, 
 $ M_{*}/L_{\rm T}^{\rm bol}$ increases of a factor of $\simeq 15$ in the first 1 {\rm Gyr} from the
burst and only of a factor of $\simeq 3$ from 5 {\rm Gyr} to 15 {\rm Gyr}.
\begin{figure}
 \epsffile{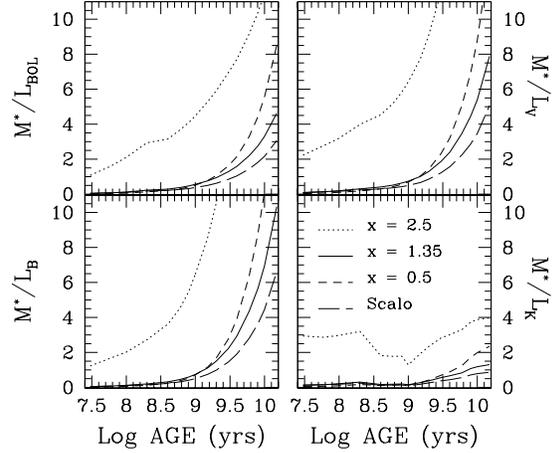}
  \caption{The dependence of stellar mass-to-light ratios on the IMF.
   \label{mlimf}}
\end{figure}
Fig. \ref{mlimf} shows the dependence of $ M_{*}/L $ on the IMF, $M_{\rm inf}$ being 
constant. At any age, the $ M_{*}/L $ ratios of a dwarf-dominated population 
($x$ = 2.5) are much higher with respect to those of a population with a 
Salpeter or flatter IMF. 
For $t \ga 1 $ {\rm Gyr} a population with a top-heavy IMF ($x$ = 0.5) has 
$ M_{*}/L $ ratios larger than those of a Salpeter SSP, because of the larger 
number of heavy remnants.
\begin{figure}
 \epsffile{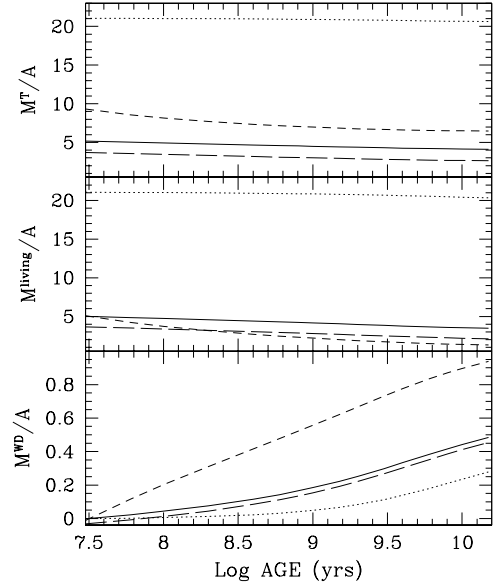}
  \caption{The evolution of the total stellar mass, the mass in living stars 
and the white dwarf component (from the top to the bottom). Solid line is for 
a Salpeter IMF, dotted line for a dwarf dominated IMF ($x$ = 2.5), dashed-line 
for a giant dominated IMF ($x$ = 0.5) and long-dashed line for a Scalo 
multislope IMF.
   \label{mass}}
\end{figure}

The $ M_{*}/L $ trend is easily understood by looking at the evolution of  
the total stellar mass and its components (Fig. \ref{mass}). 
The stellar mass of a dwarf-dominated SSP is mainly composed of living stars at
nearly any age because of the long evolutionary time-scales of low mass stars. 
A giant-dominated ($x$=0.5) SSP, on the contrary, has a larger amount of 
`remnants' from massive stars. 

We refer to a forthcoming paper for an analysis of the dependence of  
$ M_{*}/L $ on the SSP chemical composition.

\section{Comparisons with star clusters}

Synthetic broad-band colours and the luminosity contributions of evolutionary
phases are now compared with the corresponding quantities for Globular Clusters,
 the natural counterparts of SSPs models. Population size effects limit
somewhat this comparison in the case of IR colours, as each cluster contains
at most a few TP-AGB stars, hence stochastic fluctuations affect IR colours
such as $(V-K)$.  
\begin{figure*}
 \begin{minipage}{120mm}
 \epsffile{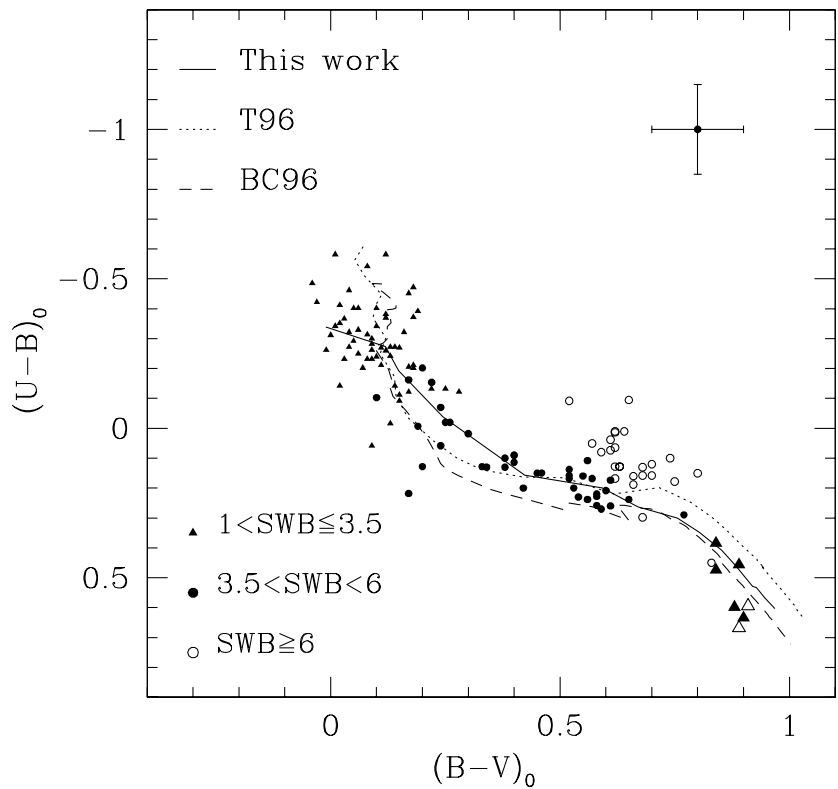}
  \caption{$(U-B)$-$(B-V)$ synthetic diagram, compared with LMC and Galactic 
globular cluster colours. LMC GCs data (from van den Bergh 1981) are
represented according to the SWB classification scheme (see the text), as 
indicated by the labels at the bottom left.
NGC 5927, NGC 6440, NGC 6624, Terzan 5, Pal 8 Galactic clusters ({\it full 
triangles}) and NGC 6528 and NGC 6553 Bulge clusters ({\it open triangles}) 
data are from Harris (1996). Reddening values have been taken from the authors
and the typical error bar is indicated. T96 and BC96 solar models are also 
plotted.
  \label{ubbv}}
   \end{minipage}
\end{figure*}

\subsection{Colour-colour diagrams}

Fig. \ref{ubbv} shows our synthetic $(U-B)$ versus $(B-V)$ relation, 
compared with LMC 
GCs and Galactic clusters data. Observational $(B-V)$ colours are corrected 
for reddening following the authors prescriptions; for Galactic clusters 
$(U-B)$ colours, the relation $E(U-B)=0.8E(B-V)$ has been adopted.
Tantalo {\rm et al.} (1996; hereafter T96) and Bruzual \& Charlot (1996, 
{\it private communication}; hereafter BC96) models are also plotted. 
\begin{figure*}
\begin{minipage}{120mm}
 \epsffile{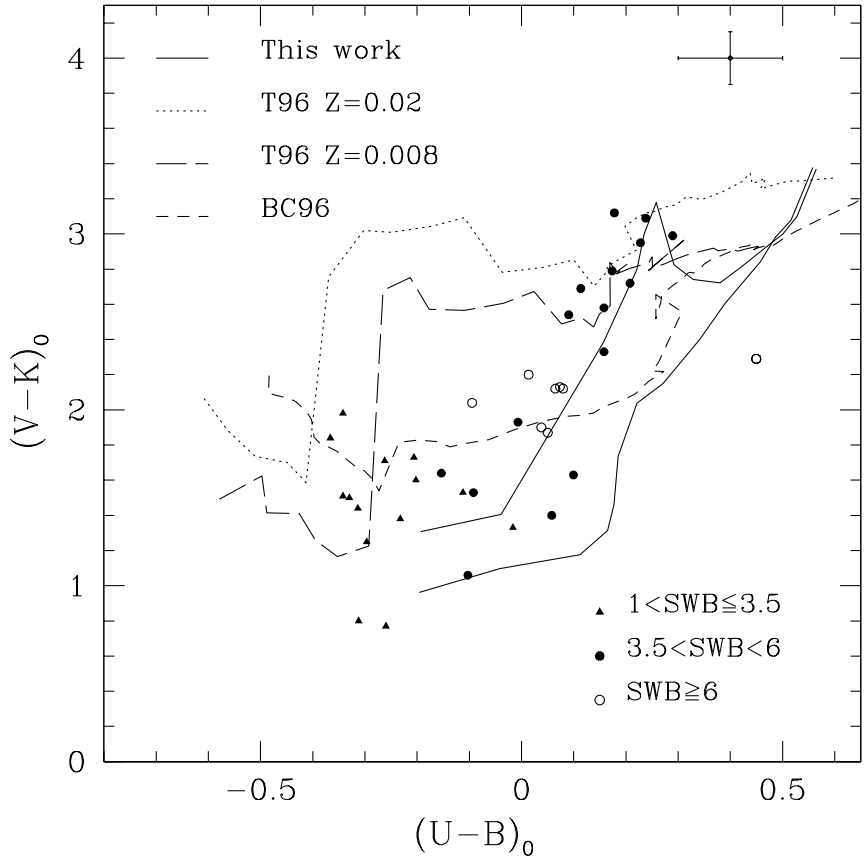}
  \caption{$(V-K)-(U-B)$ two-colour diagram for the LMC GCs of Fig. \ref{ubbv}
  with available infrared photometry (from Persson {\rm et al.} 1983);
 reddening values are taken from the authors and the typical error bar
 is indicated. As in Fig. \ref{ubbv}, clusters are represented according
  to the SWB parameter (see labels at the bottom-right).
 T96 models for two different metal content and Bruzual \& Charlot
 (1996, {\it private communication}) solar models are also plotted. 
 The lowest solid line represents our SSP model in which the TP-AGB phase 
 has been omitted.
     \label{vkub}}
   \end{minipage}
\end{figure*}
\begin{figure*}
 \begin{minipage}{120mm}	
 \epsffile{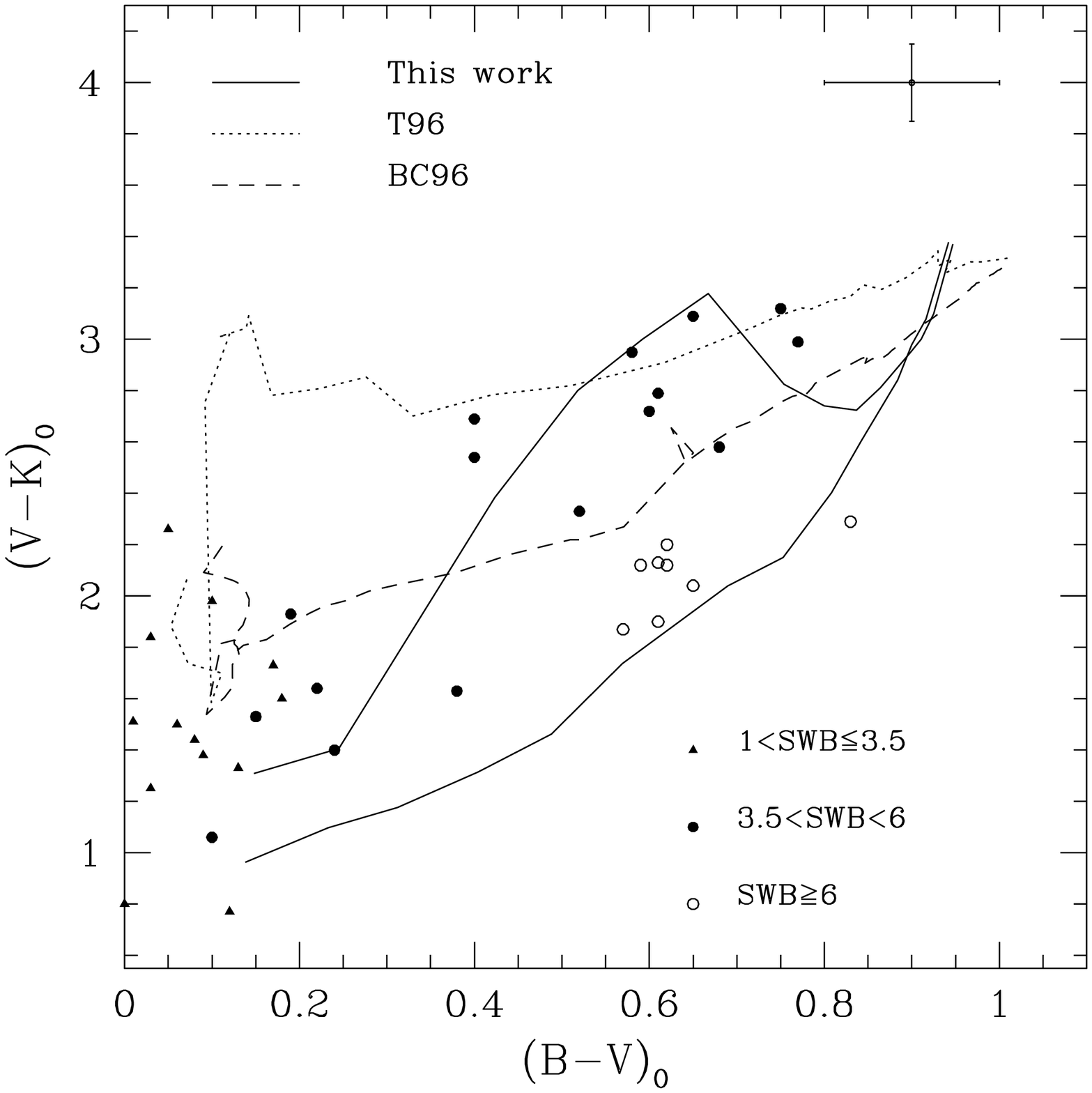}
  \caption{$(V-K)-(B-V)$ two-colour diagram for the same LMC clusters 
and models as in Fig. \ref{vkbv}.
     \label{vkbv}}
    \end{minipage}
\end{figure*}

LMC clusters are grouped according to the {\rm SWB} classification 
scheme and represented with different symbols (indicated by the labels in 
Fig. \ref{ubbv}, \ref{vkub} and \ref{vkbv}). Types 1 
to 3 correspond to young clusters (t $ \la $ 120 {\rm Myr}, little fill 
triangles); types 3.5 to 5.5 to intermediate ones (250 {\rm Myr} $ \la t \la
 $ 2 {\rm Gyr}, fill dots) and types 6 to 7 to old ones (t $ \ga $ 3 {\rm Gyr}, open circles)}. 

The discrepancy between the models and LMC data for $ (B-V) \ga 0.6$ is due to 
the increasing difference in the metal content. In fact, as already 
noticed, SWB types $>$ 6 are much more metal poor ($ Z \simeq 10^{-3} - 10^{-4}$)
 compared to earlier type clusters ($ Z \simeq 0.01$). This explains why they 
are bluer in spite of being older.

In the range $(B-V) \la $ 0.6, a good agreement is evident and the various 
models behave in much the same way. Galactic cluster colours are consistent 
with old ages, in agreement with the standard picture.

Fig. \ref{vkub} shows $(V-K)$ vs. $(U-B)$ compared with a sample of LMC GCs with 
available infrared photometry (from Persson {\rm et al.} 1983). 
 
As it is well known, intermediate age LMC clusters exhibit a wide range of  
$(V-K)$ values, from $\sim $ 1 to $\sim $ 3.5. This is the result of the 
combined effect of AGB + RGB development, also visible in a $(V-K)$ vs. 
$(B-V)$ diagram (Fig. \ref{vkbv}). The synthetic colours of our calibrated SSP model 
 well agree with the observed ones, in this age range.
T96 and BC96 models fail to reproduce the observed trend of IR colours at 
intermediate ages. Possible reasons will be discussed in the next paragraph.

\subsection{Comparing different EPS}

Analysing the discrepancies between different EPS models is not a simple job. 
Many factors contribute to produce different results, such as the stellar 
tracks used, the photometric calibrations adopted to transform the theoretical 
Hertzsprung Russel diagram into observables and the synthesis method 
itself. In this section we limit ourselves to comment the discrepancy 
between synthetic $(V-K)$ colours at intermediate ages, as shown in Fig.
\ref{vkub} and \ref{vkbv}. The comparison is made with T96 and BC96.
\begin{figure}
 \epsffile{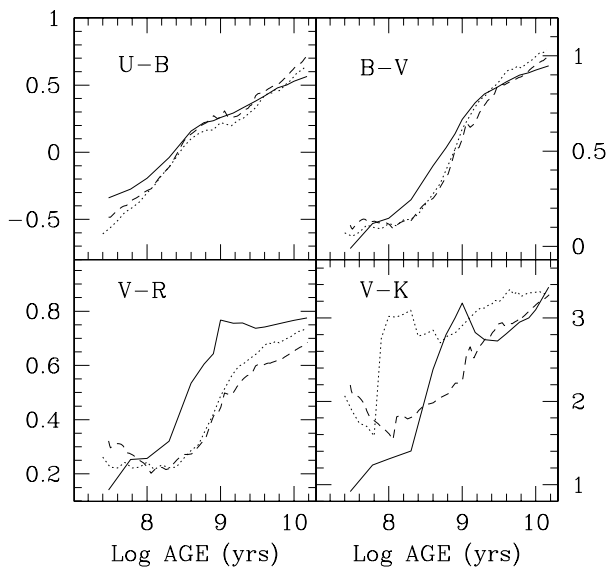}
  \caption{A comparison between different broad band colour models:
this work ({\it solid line}); T96 ({\it dotted line}); BC96
  ({\it dashed line}). All models assume a solar chemical composition, a
 Salpeter IMF, and a lower mass cutoff at 0.1 $M_{\odot}$.
   \label{colorautor}}
\end{figure}

Fig. \ref{colorautor} shows various synthetic broad-band colours as functions of age. 
The infrared indices of the three EPS models are quite different. 
The impact of AGB phase-transition at intermediate ages is visible only 
in our $(V-K)$ model. T96 anticipate its occurence, and their $(V-K)$ shows a 
jump at $t \sim 10^8$ {\rm yr}. This is in contradiction with the 
observations. For example, the LMC cluster NGC 1866 ($t \sim 10^{8} $ 
{\rm yr}) does not have a developed AGB: according to FMB90, its AGB 
bolometric contribution is only the 6 per cent of the total. 

T96 adopt the analytical prescriptions for the AGB phase as in Bressan, 
Chiosi \& Fagotto 1994 (hereafter BCF94). These are based on the 
core mass-luminosity relation from Boothroyd \& Sackmann (1988) and do not 
taken into account the results of Bl\"{o}cker \& Sch\"{o}nberner 1991 
(see 3.2), as emphasized in BCF94 (see BCF94 for more details).
   
BC96 adopt a semi-empirical receipt (described in Charlot \& Bruzual 1991) to 
determine the position
 of TP-AGB stars on the theoretical CMD diagram. In their Fig. 4, the authors 
 compare the predicted fractional luminosity contributions of the AGB with 
 FMB90 data, as we do in Fig. \ref{agbswb}, and a satisfactory agreement is evident. 
However, the integrated $(V-K)$ of BC96 does not reach the high values observed
 in the intermediate-age LMC GCs. 
This is probably due to the energetic associated with {\rm C} and {\rm M} type
 stars (see Table 2). Furthermore, Charlot, Worthey \& Bressan (1996) 
 underline the lack of carbon star spectra in Bruzual \& Charlot (1993)
 (and BC96) computations. The impact, on IR colours, of the amount of fuel 
 burned by {\rm C} or {\rm M} type stars will be discussed in sec. 5.4.2.
   
At the end, it is clear that the major source of discrepancy between all 
models is the treatment of the AGB phase, as it controls the SSP infrared 
emission at intermediate ages (cf. Fig. \ref{contrmono}). 

On the opposite, the evolution of $(U-B)$ and $(B-V)$ is very similar for all 
 sets of models. This arises from optical indices mainly depending on turn-off 
 stars (cf. Fig. \ref{contrmono}). The modest differences arise from the different input 
stellar tracks. 

\subsection{The luminosity contributions of evolutionary phases}

A fundamental and global test for SSP models is the comparison of the synthetic
 luminosity contributions of the various evolutionary stages with the same 
 quantities as measured in stellar clusters (cf. Renzini \& Fusi Pecci 1988). 
We show the results of two tests of this kind. 

The first test is performed on an intermediate age LMC GC sample 
(Ferraro {\rm et al.} 1995; hereafter F95), selected with the aim of exploring 
 the effect of AGB and RGB phase-transitions on integrated luminosities.
The observed AGB and RGB contributions to the bolometric and $K$ luminosity 
are compared with the synthetic quantities.

 As emphasized by the authors, their cluster sample is rather limited 
(12 clusters in total) and the available star sample in each cluster is
subjected to statistical fluctuations, especially in the case of AGB
 stars. AGB stars, whenever present, dominate the integrated cluster light, 
 actually driving the IR magnitudes and colours, despite their small numbers. 
 Anyway, the very short lifetimes involved ($\sim$ 3.5 {\rm Myr} for the whole
AGB, Iben \& Renzini 1983) make highly fluctuating the AGB luminosity 
contributions from one cluster to another of nearly the same age.
To limit this effect, clusters have further been grouped in age bins, 
following the procedure described in sec. 3.2. The LMC cluster NGC 1987 
($s$=35) was not taken into account, because the AGB contribution is not
properly defined in F95.

The cluster ages are derived from their $s$-parameter according to the 
relation $ \log t=0.079s+6.05 $ (Elson \& Fall 1988), as in F95. This relation 
is based on stellar evolutionary models by VandenBerg (1985),
 which adopt canonical convection (no overshooting). Our SSP models are based 
on a different set of stellar models (CCS92), which also adopt canonical
convection. The two age-turnoff luminosity relations are very similar (to
within $\sim$ 10 per cent) and therefore the Elson \& Fall $s$-age relation is
 consistent with our SSP models. 
For an $s$-age relation from overshooting models see BCF94.
\begin{figure}
 \epsffile{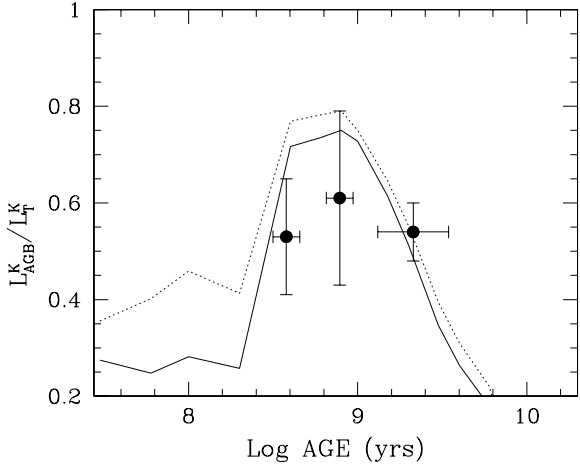}
  \caption{The AGB contribution to the total $K$ luminosity compared with LMC
GCs data from F95. The solid line represents the contribution of TP-AGB,
while the dotted one refers to the whole AGB (i.e, E-AGB plus TP-AGB). 
Error bars indicate the r.m.s. (see the text). 
 \label{f95agbk}}
\end{figure}
\begin{figure}
 \epsffile{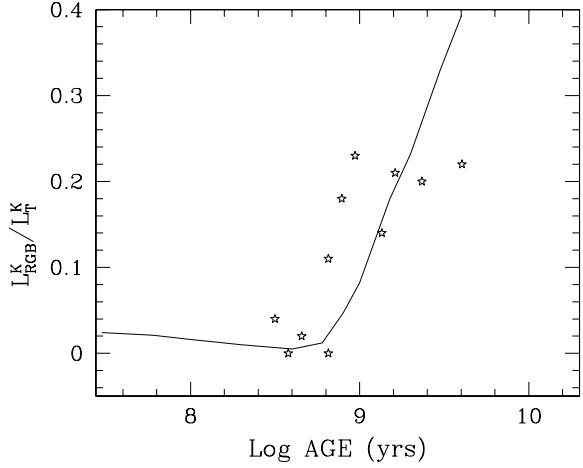}
  \caption{The RGB contribution to the total $K$ luminosity compared to
LMC GCs data from F95.
  \label{f95rgbk}}
\end{figure}
\begin{figure}
 \epsffile{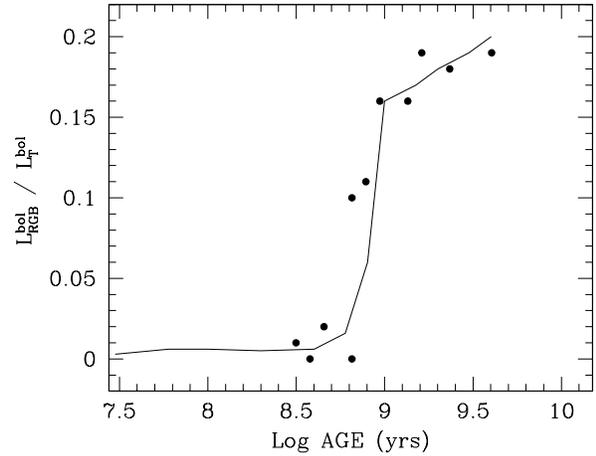}
  \caption{The RGB bolometric contribution compared to F95 data. The synthetic 
$L_{\rm RGB}^{\rm bol}/L_{\rm T}^{\rm bol}$ takes into account that F95 sample is limited to 
 2.3 {\rm mag.} fainter than the TIP of the RGB.
  \label{f95rgbbol}}
\end{figure}

The results for the AGB phase are shown in Fig. \ref{f95agbk}. The solide line 
represents the synthetic TP-AGB contribution to the total $K$-light, and the 
dotted line refers to the whole AGB (that is, E-AGB plus TP-AGB). The error 
bars indicate the r.m.s. 

The comparison shown in Fig. \ref{f95agbk} suggests an agreement between the observed 
AGB phase transition and the synthetic luminosity contribution of the AGB 
phase, that, we recall, has been calibrated on the LMC GCs, as described in 
sec. 3.2. 
     
Fig. \ref{f95rgbk} and \ref{f95rgbbol} show the RGB contribution to the $K$ and bolometric 
luminosities, compared with the same quantities from F95.
Our model simulates the onset and early development of RGB phase
transition very nicely.

For old ages ($t\sim$ 15 {\rm Gyr}), the test is made with the $\sim
Z_{\odot}$ Bulge clusters NGC 6528 and NGC 6553, recently observed with HST 
(Ortolani {\rm et al.} 1995).
 
The empirical contributions of evolutionary stages to the V-luminosity 
(Lanteri Cravet {\rm et al.} 1997 and Lanteri Cravet, {\it private 
communication}) are compared with the corresponding theoretical quantities in 
Table 5. To allow a 
more significative comparison, the synthetic MS $V$-luminosity has been 
truncated at the completeness limits of the samples ($\sim 1.29$ mag. 
below the TO for NGC 6553, for which $V_{\rm lim}^{6553}=21.5$, 
Ortolani, {\it private communication}; $\sim 2.95$ mag. below the TO, for NGC 
6528, for which $V_{\rm lim}^{6528}=23.6$, Marconi, 
{\it private communication}).
\begin{table*}
 \centering
  \begin{minipage}{110mm}
  \caption{The contributions of evolutionary phases to the total 
  $V$-luminosity (for a Salpeter IMF) compared with the observational data of 
the Bulge clusters NGC 6528 and NGC 6553 (Lanteri Cravet {\rm et al} 1997 and
lanteri Cravet {\it private communications})}.
   \begin{tabular}{@{}ccccccc}
& & & & & & \\
Phase $j$ & NGC 6528 & 10 {\rm Gyr} & 15 {\rm Gyr} & NGC 6553 & 10 {\rm Gyr} & 
15 {\rm Gyr} \\
& & & & & & \\
MS + SGB & 0.492$\pm 0.006$ & 0.457 & 0.483 & 0.421 $\pm 0.04$ & 0.402 & 0.432 \\
RGB & 0.271 $\pm 0.01$ & 0.274 & 0.273 & 0.320 $\pm 0.01$ & 0.302 & 0.300 \\
HB & 0.154 $\pm 0.02$ & 0.181 & 0.163 & 0.180 $\pm 0.01$ & 0.199 & 0.179 \\
AGB & 0.083 $\pm 0.02$ & 0.087 & 0.081 & 0.08 $\pm 0.01$ & 0.096 & 0.089  \\ 
   \end{tabular}
 \end{minipage}
\end{table*}
\begin{table*}
 \centering
  \begin{minipage}{125mm}
  \caption{The IMF influence on evolutionary phase luminosity contributions 
  of Table 6}
   \begin{tabular}{@{}ccccccccc}
& & & & & & & & \\
&\multicolumn{4}{c}{ $ \Delta {V^{\rm lim}_{TO}}$ = 2.95}
&\multicolumn{4}{c}{$ \Delta {V^{\rm lim}_{TO}}$ = 1.29} \\
&\multicolumn{2}{c}{$x$ = 2.5} &\multicolumn{2}{c}{$x$ = 0.5} 
&\multicolumn{2}{c}{$x$ = 2.5} &\multicolumn{2}{c}{$x$ = 0.5 }\\
& & & & & & & & \\
Phase $j$ & 10 {\rm Gyr} & 15 {\rm Gyr} & 10 {\rm Gyr} & 15 {\rm Gyr} & 10 {\rm
Gyr} & 15 {\rm Gyr} & 10 {\rm Gyr} & 15 {\rm Gyr} \\
& & & & & & & & \\
MS + SGB & 0.476 & 0.502 & 0.444 & 0.471 & 0.410 & 0.441 & 0.397 & 0.425 \\
RGB & 0.265 & 0.263 & 0.281 & 0.280 & 0.298 & 0.295 & 0.305 & 0.303 \\
HB & 0.175 & 0.157 & 0.185 & 0.167 & 0.197 & 0.176 & 0.201 & 0.181 \\
AGB & 0.084 & 0.078 & 0.089 & 0.083 & 0.095 & 0.088 & 0.093 & 0.090 \\ 
   \end{tabular}
 \end{minipage}
\end{table*}
The theoretical contributions in Table 5 are fairly insensitive to the adopted
 IMF (see Table 6). In fact, down to the $V_{\rm lim}$ values quoted
above, the stellar counts sample a small mass-range (for the 15 {\rm Gyr}
 SSP we have: $ 0.82 \la M \leq 0.938 $ for $ \Delta {V^{\rm lim}_{TO}} $ = 
 1.29, $ 0.7 \la M \leq 0.938 $ for $ \Delta {V^{\rm lim}_{TO}} $ = 2.95).
The agreement between predicted and observed contributions is quite 
satisfactory, and an old age ($\simeq 15$ {\rm Gyr}) seems to be favoured, in 
accordance with Ortolani {\rm et al.} (1995) conclusions.

\subsection{Two experiments}

Thanks to the modular structure of the code it is possible to explore the 
impact of changing one or more ingredients and/or assumptions, on integrated 
photometric properties. We show here the results of two tests: the 
simulation of an `intermediate' morphology for the Horizontal Branch at late 
ages, and the effect of a different contribution of $M$ and carbon stars at 
intermediate ages. 

\subsubsection{Effect of a Blue HB}

The location of the Helium-Burning (HB) stars of Globular Clusters on a HR 
diagram displays different morphologies, called `red-HB' (RHB), `blue-HB'
(BHB) or `intermediate HB' (IHB), depending on the observed extension 
towards blue. As well known, the HB morphology shows a correlation with
cluster metallicity (the 'first parameter'), but exceptions are so numerous
that a 'second parameter' is required. The nature of this second parameter 
has not been firmly established, but commonly considered candidates include
age and/or the cluster dynamical state (see Fusi Pecci {\rm et al.} 1993).
On average, metal-poor clusters have blue HB stars, while metal-rich ones 
show red HB. However, the presence of BHB stars in old, metal rich SSPs is
theoretically plausible (see Greggio \& Renzini 1990) and now observationally
confirmed for some metal rich clusters in the galactic Bulge 
(Rich {\rm et al.} 1997). 
In this section we explore the effect on integrated colours of adding BHB
stars on the 15 {\rm Gyr} SSP, while keeping constant the HB fuel consumption.
In this way, we simulate an IHB, as the red component is maintained.
For this test we modify the $(T_{\rm e},g)_{\rm ij}$ matrix at $t_{\rm i}$=15 
{\rm Gyr}, $j$=HB, extending the temperature distribution to 
$\log T_{\rm e}\simeq 4$; a flat stellar distribution is assumed. 
The surface gravity for each effective temperature is derived assuming a 
constant luminosity.
\begin{figure}
 \epsffile{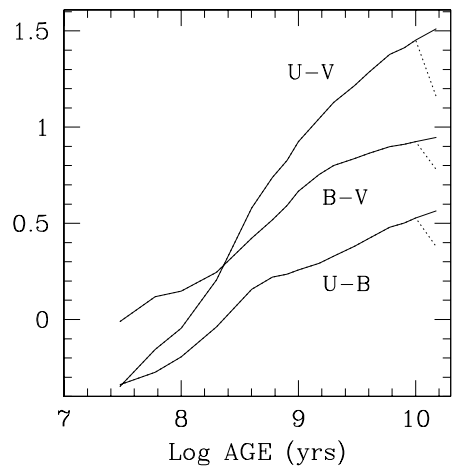}
  \caption{The effect on integrated colours of adding a BHB to the 15 
{\rm Gyr} SSP. Dotted lines are SSP models in which an intermediate HB 
morphology is simulated.
    \label{bhb}}
\end{figure}

The result is shown in Fig. \ref{bhb}. As expected, the $(U-V)$ colour is the 
most influenced by the introduction of blue HB stars. 

\subsubsection{Varying the fraction of carbon stars}

The initial metal content $Z$ of a stellar population plays a key role in 
determining the production and the characteristics of carbon stars during the 
AGB phase (RV81). By reducing $Z$, the abundance of oxygen in 
the envelope is lower, thus a lower amount of carbon has to be dredge-up in 
order to achieve {\rm C/O} $ > $ 1 and create carbon stars. 
Therefore, a metal poor stellar population is expected to display much more 
carbon stars than a metal rich one of the same age. 
Passing from Z $\sim $ 0.008 (appropriate for LMC) to $Z_{\sun}$, the 
fraction of carbon stars is reduced by a factor of $\sim$ 2. To estimate this 
effect on integrated colours, we have reduced by a factor of 2 the amount of 
fuel burned by {\rm C} stars (see Table 2), assigning it to {\rm M} stars. 
\begin{figure}
 \epsffile{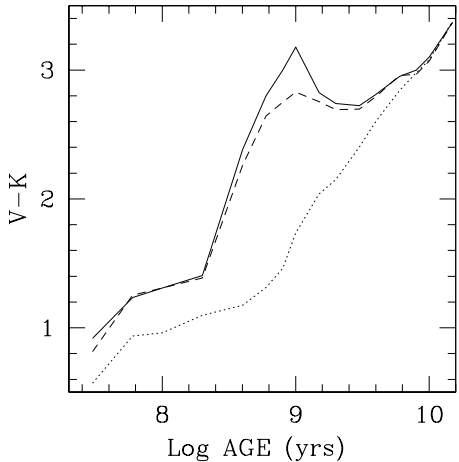}
 \caption{The influence of carbon stars populating the TP-AGB 
 phase on integrated $(V-K)$. Solid and dotted lines are the models of Fig.
 \ref{colorage} (including or not the TP-AGB contribution respectively); 
 dashed line is a model in which the amount of TP-AGB fuel for {\rm C} stars 
 has been reduced by a factor of two.
  \label{vksolar}}
\end{figure}

Fig. \ref{vksolar} shows the result: the AGB phase-transition is still 
evident in the integrated $(V-K)$ at intermediate ages, but the colour 
maximum value is reduced by $\sim $ 0.4 mag.  

\section{Conclusions}

In this paper we have used the Fuel Consumption Theorem approach to 
compute evolutionary synthesis models (EPS) for Simple Stellar Populations 
(SSPs). The computational code has a modular structure, in which the main 
ingredients are kept as separate entries. This allows to investigate their 
impact on the results and to better assess the uncertainties. 
The TP-AGB contribution has been calibrated on LMC globular Clusters.
A grid of SSP models, having a solar chemical composition and covering a wide 
age-range (30 {\rm Myr} - 15 {\rm Gyr}), has been constructed and submitted to
 several observational tests. 

Synthetic broad-band colours, compared with LMC GCs data, show a fair
agreement. In particular, the observed rise in $(V-K)$ colour at intermediate 
ages is well reproduced, as a consequence of the adopted calibration of the 
fuel consumption along the AGB phase. Comparing our results with others in the 
literature (Tantalo {\rm et al.} 1996 and Bruzual \& Charlot (1996; 
{\it private communication}), we show 
that only a correct calibration of the AGB energetics allows to match 
simultaneously the observed optical and IR properties of intermediate-age 
clusters.

A complete test of the luminosity contributions by the various synthetic 
evolutionary phases has been performed on the old Bulge clusters NGC 6553 and 
NGC 6528. The result is very satisfactory and an old age (t $\simeq$ 15 
{\rm Gyr}) for these clusters is favoured. 

Having so tested the procedure, these models along with similar ones for
 other chemical compositions now being computed, will be used for a variety of
 astrophysical applications.
   
\section*{Acknowledgments}

The author is indebted with Alvio Renzini and Laura Greggio for their guidance 
 and for their critical reading of this paper. The author also
gratefully acknowledges Luca Ciotti, Francesca Lanteri Cravet and Sergio
Ortolani for many stimulating discussions. It is a pleasure to thank Gustavo 
Bruzual for kindly providing his SSP models in advance of publication and 
Oscar Straniero for his theoretical tracks in a computer readable form.
It is also a pleasure to thank the European Southern Observatory and the 
Universit\"{a}t Sternwarte-M\"{u}nchen for their kind and generous hospitality 
during part of my research work. Finally, a special acknowledgement to the 
anonymous referee, whose helpful comments have greatly improved the 
presentation.

\appendix

\section[]{Tables}

The SSP models listed in the following tables refer to the TP-AGB
fuel calibration described in Sec. 3.2. 

\begin{table*}
  \centering
 \begin{minipage}{140mm}
  \caption{Integrated colours and stellar mass-to-light ratios as functions 
 of age and IMF slope}
  \begin{tabular}{@{}cccccccccccc}
 & & & & & & & & & & & \\
AGE & $(U-V)$ & $(U-B)$ & $(B-V)$ & $(V-R)$ & $(V-K)$ & $ M^*/L_{bol} $ 
&  $ M^*/L_{U} $ & $ M^*/L_{B} $ & $ M^*/L_{V} $ & 
 $ M^*/L_{R} $ &  $ M^*/L_{K} $ \\  
 & & & & & & & & & & & \\
\multicolumn{12}{c}{$x$ = 0.5} \\ 
& & & & & & & & & & & \\
0.03 &  -0.305  &   -0.314 &     0.009 &     0.163 &      1.018 &
0.013  &  0.010 &  0.015 &  0.027 &   0.038  &  0.040 \\
0.06  &  -0.113   &  -0.264  &    0.151  &    0.284   &   1.345 &
  0.032 & 0.026  &  0.037  & 0.059  &  0.073 &   0.063 \\
0.1  &  -0.019 &    -0.190 &     0.172  &    0.281 &     1.405 &
  0.061 &    0.046 &   0.062 &    0.096 &    0.119 &    0.097 \\
0.2 &   0.236   &  -0.033 &     0.269  &     0.342  &    1.481 & 
  0.134 &    0.111 &    0.128 &    0.183 &   0.216 &    0.173 \\
 0.4 &    0.626 &     0.177  &    0.449 &     0.562  &     2.468 &
 0.199 &    0.253 &    0.243 &   0.292 &    0.281 &    0.111 \\
 0.6  &   0.786 &     0.245  &    0.540 &     0.626 &     2.868 &
 0.300 &    0.419 &    0.377 &   0.417 &    0.378 &    0.139 \\
  0.8 &  0.871 &  0.259  &    0.611 &  0.660  &    3.056 &
     0.400 &    0.595 &    0.528 &    0.548 &    0.481 &    0.168 \\
 1  &  0.968  &    0.283  &     0.685  &     0.786  &     3.237 &
  0.506 &  0.815 &    0.708 &    0.685 &    0.536 &    0.128 \\
 1.5  & 1.078 &     0.312  &    0.767 &     0.766  &    2.857 &
  0.774 &    1.369 &    1.158 &    1.040 &    0.829 &    0.276 \\
 2  &  1.154  &    0.344 &   0.809  &   0.762 &   2.760 &
  1.048 &   2.046 &    1.679 &    1.450 &    1.160 &    0.422 \\
 3  & 1.229  &    0.388   &   0.840 &     0.734  &    2.723 &
   1.609 &    3.520 &    2.774 &    2.327 &    1.911 &    0.701 \\
 4  &  1.290   &   0.425  &    0.865  &    0.738   &   2.803 &
   2.150 &    5.141 &    3.919 &    3.216 &   2.638 &    0.899 \\
 6  & 1.375  &    0.480  &    0.896  &    0.741 &     2.940 &
   3.275 &    8.773 &    6.356 &    5.068 &    4.137 &    1.250 \\
 8  &  1.404   &   0.499  &    0.905   &   0.720  &    2.960 &
  4.416 &   12.323 &    8.773 &    6.935 &    5.774 &    1.716 \\
 10  &   1.440  &    0.523  &    0.917 &     0.728 &     3.035 &
 5.557 &   16.122 &   11.221 &    8.780 &    7.251 &    1.983 \\
 15  & 1.495   &   0.559    &  0.936  &    0.746  &    3.343 & 
  8.496  & 26.606 &   17.959 &   13.785 &   11.205 &    2.347 \\
 & & & & & & & & & & & \\
\multicolumn{12}{c}{$x$ = 1.35} \\ 
 & & & & & & & & & & & \\
        0.03 &   -0.349  &   -0.339 &    -0.011 &     0.141 &     0.918
& 0.036 &  0.028 & 0.043 & 0.079 & 0.112 & 0.125 \\

        0.06 & -0.155    &  -0.274  &    0.119  &    0.253  &    1.235 &
       0.074 &   0.056 &   0.082 &   0.133 &   0.171 &    0.158        \\

       0.1 & -0.046    & -0.194   &   0.148   &   0.257  &    1.308 &
 0.120 &    0.087 &    0.117 &   0.186 &   0.236 & 0.206 \\
       0.2 & 0.206  &   -0.039   &   0.245  &    0.321 &     1.405 &
 0.220 &  0.171 & 0.201 &  0.292 &    0.351 &    0.297 \\
       0.4 & 0.580  &    0.157  &    0.423  &    0.535 &     2.384 &
 0.279 & 0.330 &  0.322 &    0.397 &    0.391 &   0.163\\
       0.6 & 0.739  &    0.221  &    0.518  &    0.605 &  2.800
 & 0.377 & 0.493 & 0.454 &    0.512 &    0.474 &    0.182 \\
       0.8 & 0.827  &    0.235  &    0.592  &    0.643  &    3.000
& 0.465 & 0.653 & 0.593 &   0.625 &    0.560 &    0.202 \\
      1 & 0.924  &    0.258  &    0.667  &    0.767  &    3.178 &
 0.552 & 0.843 &  0.750 & 0.740 & 0.588 & 0.146 \\
      1.5 & 1.045  &    0.292  &    0.754  &    0.756 &     2.824 &
 0.753 &  1.289  &  1.111  &  1.009 &    0.812 &    0.277  \\
      2 & 1.129  &    0.329  &    0.800  &    0.757 &     2.741 &
 0.941 &  1.797 &   1.496 &  1.303 &  1.047  &  0.386  \\
      3 & 1.216  &    0.380  &    0.837  &    0.737 &     2.723 &
 1.289 &  2.810 &    2.232 &    1.879 &    1.539 &  0.566 \\
      4 & 1.285  &    0.421  &    0.864  &    0.742 &     2.812 &
 1.591 &    3.836 &    2.940 &    2.410 &    1.965 &    0.669 \\
      6 & 1.378  &    0.479  &    0.899  &    0.753 &     2.950 &
 2.187 &    5.995 &  4.347 &    3.460 &  2.791 &    0.840 \\
      8 & 1.413  &    0.501  &    0.911  &    0.760 &     3.000 &
 2.753 &    7.969 &   5.658 &    4.447 &    3.641 &    1.069 \\
     10 & 1.452  &    0.528  &     0.925 &    0.766 &      3.100 &
 3.295 &   10.012 &    6.948 &   5.389 &   4.364 & 1.181 \\
     15 & 1.512  &    0.565  &    0.947  &    0.775 &     3.370 &
 4.616 &  15.446 &  10.332 &  7.867 &    6.227 &    1.306  \\
 & & & & & & & & & & & \\
\multicolumn{12}{c}{$x$ = 2.5} \\ 
 & & & & & & & & & & & \\
        0.03 & -0.322 &  -0.329 &  0.007 & 0.165 &  1.116
      & 1.076  &  0.806  &  1.231 &   2.226 &    3.087 &  2.946 \\  
        0.06 & -0.141 &  -0.252 &  0.111 & 0.256 & 1.387
     & 1.596 &  1.187 &    1.688 &   2.773 &    3.535 &  2.859 \\
       0.1   & -0.024 & -0.173  & 0.149 & 0.276 & 1.497
& 2.095 & 1.521 &  2.011 &  3.190 &  3.994 &  2.971 \\
       0.2  & 0.209  & -0.034   & 0.243 & 0.345 & 1.657
& 2.936 &  2.349 &  2.733 &  3.974 & 4.669 & 3.194 \\
       0.4  & 0.544  &  0.133  & 0.410  & 0.534  & 2.404
& 3.153 & 3.633 & 3.623 & 4.517 & 4.459 & 1.825 \\      
       0.6  & 0.700  & 0.192 & 0.508 & 0.610  & 2.820
& 3.676 & 4.818 & 4.551 & 5.184  & 4.772 & 1.777 \\       
       0.8  & 0.793 & 0.208  & 0.585  & 0.655  & 3.054
 & 4.089 & 5.885 & 5.475 & 5.815 & 5.135 & 1.791 \\      
        1  & 0.891  & 0.232  & 0.659  & 0.768  & 3.165
 & 4.471 & 7.103 & 6.470 & 6.416  & 5.105 & 1.286 \\       
      1.5  & 1.028 & 0.275  & 0.753  & 0.777 & 2.921
 & 5.264 & 9.753 & 8.533 & 7.758 & 6.121 & 1.947 \\     
      2  & 1.125 & 0.320  & 0.806  & 0.791  & 2.897
 & 5.902 & 12.543 & 10.532 & 9.125 & 7.113 & 2.342 \\     
      3 & 1.234  & 0.381  & 0.853  & 0.797  & 2.945
& 6.951 & 17.574 & 13.943 & 11.569 & 9.015 & 2.840 \\
      4 & 1.316  & 0.429  & 0.887  & 0.807  & 3.052
& 7.716 & 22.223 & 16.875 & 13.566 & 10.415 & 3.018 \\     
      6 & 1.425  & 0.495  & 0.930  & 0.835  & 3.219
& 9.148 & 31.288 & 22.368 & 17.289 & 12.921 & 3.297 \\      
      8 & 1.473  & 0.524  & 0.949  & 0.850  & 3.271
& 10.373 & 38.880 & 27.038 & 20.520 & 15.310 & 3.729 \\      
     10 & 1.521  & 0.553  & 0.967  & 0.860  & 3.376
& 11.461 & 46.436 & 31.452 & 23.482 & 17.163 & 3.874 \\     
     15 & 1.598  & 0.598  & 1.001  & 0.909  & 3.646
& 13.808 & 65.494 & 42.537 & 30.837 & 21.535 & 3.965 \\     
  \end{tabular}
 \end{minipage}
\end{table*}
\begin{table*}
  \centering
  \caption{$A/L_{\rm T}$ and the bolometric correction factors 
  $ B_{\rm c}^{\lambda}$, as functions of age and IMF slope}
   \begin{tabular}{@{}ccccccc}

 & & & & & & \\
AGE (\rm Gyr) & $A/L_{\rm T}$ & $ B_{\rm c}^{\rm U}$ & $ B_{\rm c}^{\rm B}$ & $ B_{\rm c}^{\rm
V}$ & $ B_{\rm c}^{\rm R}$ & $ B_{\rm c}^{\rm K}$ \\ 
 & & & & & & \\
\multicolumn{7}{c}{$x$ = 0.5} \\ 
 & & & & & & \\
        0.03   & 0.001   &   0.778    &  1.171   &   2.114  &    2.938   &  
       3.062 \\
       0.06 &        0.004  &    0.798  &    1.147  &    1.816  &  2.256 &
       1.946 \\
       0.1   &  0.008 & 0.752 &  1.010 & 1.569 &  1.955  & 1.591 \\
  0.2  & 0.017  & 0.826   &   0.959  & 1.363 & 1.606 & 1.289 \\
  0.4  & 0.027  & 1.271  &  1.217  & 1.464 & 1.409 & 0.557 \\  
  0.6 &   0.042 & 1.397  & 1.256 & 1.390  & 1.260  & 0.463 \\    
  0.8  &  0.057  & 1.488  & 1.321 & 1.368 &  1.203 &  0.421 \\
  1    &  0.073  & 1.609 & 1.398  & 1.354 & 1.060 & 0.254 \\
  1.5 &   0.113  & 1.769  & 1.497  & 1.345 & 1.072  & 0.358 \\
  2  & 0.155  & 1.952 & 1.602 & 1.384 & 1.107 & 0.403 \\
  3  &  0.241 & 2.187  &  1.723   &   1.446  &  1.187 & 0.436 \\
  4  &   0.325 & 2.391 & 1.823 & 1.496  & 1.227 & 0.419 \\
  6  &  0.500 & 2.679  & 1.941 & 1.548  & 1.263 & 0.382 \\
  8  &  0.678  & 2.790 & 1.987 & 1.570 & 1.307 & 0.389 \\
  10  &  0.855  & 2.901 & 2.019 & 1.580 & 1.305 & 0.357 \\
  15  &   1.314 &  3.132  &  2.114 & 1.623 & 1.319 & 0.276 \\
 & & & & & & \\
\multicolumn{7}{c}{$x$ = 1.35} \\ 
& & & & & & \\
  0.03  &       0.007 &     0.765 &     1.177 &     2.164 &  3.066  &  3.434 \\
  0.06 &    0.015  &     0.763  &    1.107 &     1.805  &    2.308  &  2.141 \\
  0.1 &   0.024 &   0.722 &   0.973 &    1.546 &  1.969 &     1.714 \\
  0.2  & 0.046  & 0.786  & 0.918 & 1.333  & 1.601 &  1.351 \\
  0.4   & 0.060  & 1.180 &  1.151  & 1.419  & 1.399 & 0.584 \\
  0.6  &  0.082  & 1.309 & 1.204 & 1.359  & 1.256 &  0.482 \\
  0.8  & 0.102  & 1.406  & 1.277  & 1.346  & 1.202 & 0.436 \\
   1  &  0.123  & 1.529  & 1.360 &  1.339  & 1.067 & 0.265 \\
  1.5  &  0.170 &  1.712 & 1.474  & 1.340 & 1.078  & 0.368 \\
   2  &   0.215 & 1.910  & 1.590  & 1.385  & 1.113  & 0.410 \\
   3  &   0.299  &  2.179 & 1.731 & 1.457 & 1.193 & 0.439 \\
   4  &  0.373 & 2.410  & 1.846 & 1.515 & 1.235 & 0.420 \\
   6  &   0.520 & 2.741 & 1.988 & 1.581 & 1.276 & 0.383 \\
   8  & 0.659 & 2.895 & 2.056 & 1.616 & 1.323 & 0.389 \\
  10  &   0.793 & 3.039 & 2.109 & 1.636 & 1.325 & 0.358 \\
  15 & 1.120  & 3.345 & 2.238 & 1.704 & 1.349 & 0.283 \\
 & & & & & & \\
\multicolumn{7}{c}{$x$ = 2.5} \\ 
 & & & & & & \\
  0.03  &  0.051 & 0.750 & 1.144 & 2.069 & 2.869  & 2.738  \\
  0.06 &   0.076 &  0.744  &  1.058 & 1.737 & 2.214 & 1.791\\
  0.1 & 0.100  &  0.726  & 0.960  & 1.523  & 1.907  & 1.418 \\
  0.2  & 0.140 &  0.800  & 0.931  &  1.354  & 1.590 &  1.088\\
  0.4   & 0.150  & 1.152 & 1.149 & 1.433 & 1.414 & 0.579 \\
  0.6  & 0.176 &  1.310 & 1.238 & 1.410 &  1.298 & 0.483\\
  0.8  & 0.196 & 1.439 & 1.339 & 1.422 & 1.256 & 0.438 \\
   1  & 0.214 & 1.589 & 1.447 & 1.435 & 1.142 & 0.288 \\
  1.5  & 0.252  &  1.853 & 1.621 & 1.474 & 1.163 & 0.370\\
   2  & 0.283  & 2.125 & 1.785 & 1.546 & 1.205 & 0.397 \\
   3  & 0.334 & 2.528  &  2.006 & 1.664 & 1.297 & 0.409 \\
   4  & 0.372 &  2.880  & 2.187 & 1.758 & 1.350 & 0.391\\
   6  & 0.442 &  3.420 & 2.445 & 1.890 & 1.412 & 0.360 \\
   8  & 0.502 & 3.748 & 2.607 & 1.978 & 1.476 & 0.360\\
  10  & 0.555 & 4.052 & 2.744 & 2.049 & 1.498 & 0.338  \\
  15 & 0.669 & 4.743 & 3.080 & 2.233 & 1.559 & 0.287 \\

  \end{tabular}
\end{table*}

\end{document}